\documentclass[11pt]{article}
\newcommand{\Comment}[1]{{}}
\usepackage{amsfonts,amsthm,amsmath,amssymb,slashed}
\usepackage[textwidth = 430 pt, textheight = 630 pt]{geometry}
\usepackage{color}

\Comment{\usepackage{color}
\definecolor{MyDarkBlue}{rgb}{0.15,0.15,0.45}
\usepackage[linktocpage=true]{hyperref}
\hypersetup{
colorlinks=true,
citecolor=MyDarkBlue,
linkcolor=MyDarkBlue,
urlcolor=MyDarkBlue,
pdfauthor={Horatiu Nastase and Jacob Sonnenschein},
pdftitle={The title},
pdfsubject={hep-th}
}

\usepackage[numbers,sort&compress]{natbib}
\usepackage{hypernat}}
\usepackage{graphicx}
\usepackage{cite}
\usepackage{url}

\newcommand\ignore[1]{}
\def\one{{\,\hbox{1\kern-.8mm l}}}

\def\Tr{{\rm Tr\, }}

\def\a{\alpha}\def\b{\beta}

\def\d{\partial}
\def\dag{\dagger}

\def\Tr{\mathop{\rm Tr}\nolimits}

\newcommand{\Cset}{{\,\,{{{^{_{\pmb{\mid}}}}\kern-.45em{\mathrm C}}}}}

\newcommand{\be}{\begin{equation}}
\newcommand{\bea}{\begin{eqnarray}}

\newcommand{\ee}{\end{equation}}
\newcommand{\eea}{\end{eqnarray}}

\newcommand{\expect}[1]{ \langle #1 \rangle} % expectation value
\newcommand{\ctau}{\check{\tau}}
\usepackage[export]{adjustbox} %for positioning images
\usepackage[]{placeins} %defines a command \FloatBarrier and the [section] option puts it before every new section

\begin{document}

\renewcommand{\thefootnote}{\fnsymbol{footnote}}

\makeatletter
\@addtoreset{equation}{section}
\makeatother
\renewcommand{\theequation}{\thesection.\arabic{equation}}

\rightline{}
\rightline{}
%   \vspace{1.8truecm}

%\begin{flushright}
% preprint nrs.
%\end{flushright}

%\vspace{10pt}

%\begin{document}
\begin{center}
{\LARGE \bf{\sc Krylov complexity from a simple quantum mechanical model for a radiating black hole}} 
\end{center} 
 \vspace{1truecm}
\thispagestyle{empty} \centerline{
%{\large \bf {\sc Cameron Beetar${}^{a},$}}\footnote{E-mail address: \Comment{\href{mailto:cameron.beetar@gmail.com}}
%{\tt cameron.beetar@gmail.com}}
{\large \bf {\sc Eric L Graef${}^{a},$}}\footnote{E-mail address: \Comment{\href{mailto:eric.graef@unesp.br}}
{\tt eric.graef@unesp.br}}
{\large \bf {\sc Jeff Murugan${}^{b},$}}\footnote{E-mail address: \Comment{\href{mailto:jeff.murugan@uct.ac.za}}
{\tt jeff.murugan@uct.ac.za}} 
{\large \bf {\sc Horatiu Nastase${}^{a}$}}\footnote{E-mail address: \Comment{\href{mailto:horatiu.nastase@unesp.br}}
{\tt horatiu.nastase@unesp.br}}}
\centerline{
{\bf{\sc and}}
{\large \bf {\sc Hendrik J R  Van Zyl${}^{b}$}}\footnote{E-mail address: \Comment{\href{mailto:hjrvanzyl@gmail.com}}{\tt hjrvanzyl@gmail.com}}
                                                        }

\vspace{.5cm}

\centerline{{\it ${}^a$Instituto de F\'{i}sica Te\'{o}rica, UNESP-Universidade Estadual Paulista}} 
\centerline{{\it R. Dr. Bento T. Ferraz 271, Bl. II, Sao Paulo 01140-070, SP, Brazil}}
\vspace{.3cm}
\centerline{{\it ${}^b$The Laboratory for Quantum Gravity and Strings,}}
\centerline{{\it Department of Mathematics and Applied Mathematics, }} 
\centerline{{\it University of Cape Town, Cape Town, South Africa}}
%\vspace{.3cm}

\vspace{1truecm}

%%%%%%%%%%%%%%%%%
\thispagestyle{empty}

\centerline{\sc Abstract}

\vspace{.4truecm}

\begin{center}
\begin{minipage}[c]{380pt}
{\noindent 
We investigate Krylov complexity in a simple quantum mechanical model describing a black hole coupled to its radiation. The model is constructed as a simplified ``mini-BMN" matrix system inspired by a recent proposal of Maldacena. Our aim is not to reproduce the full dynamics of the BMN matrix model, but rather to isolate a tractable setting in which the information-theoretic behaviour of a radiating black hole can be studied explicitly. We analyze both the early- and late-time behaviour of Krylov complexity and the associated Krylov entropy. At early times, perturbative and numerical analyses reveal the expected growth characteristic of chaotic quantum dynamics. At late times, however, the dynamics saturates to a plateau, consistent with equilibration between the black hole and its radiation and with general expectations from finite-entropy quantum systems. We argue that this plateau behaviour admits a semiclassical interpretation in terms of Euclidean instanton contributions in an effective path-integral.
The toy model studied here offers a controlled framework in which these features can be investigated analytically and numerically.
}
\end{minipage}
\end{center}

\vspace{.5cm}

\setcounter{page}{0}
\setcounter{tocdepth}{2}

\newpage

\tableofcontents
\renewcommand{\thefootnote}{\arabic{footnote}}
\setcounter{footnote}{0}

\linespread{1.1}
\parskip 4pt

%{}~
%{}~

%---------------------------------------------------------

%%%%%%%%%%%%%%%%%%%%%%%%%%%%%%%%%%%%%%%%%%%%%%%%%%%%%%%%%%%%%%%%%%%%%%%%%%%%%%%%%%%%%%%%

\section{Introduction}
Understanding how quantum information spreads in strongly interacting many-body systems has become one of the central problems at the interface of quantum gravity, quantum chaos, and quantum information theory. Black holes occupy a particularly important role in this context, since they are widely believed to represent the fastest scramblers allowed by quantum mechanics while at the same time exhibiting thermodynamic behaviour characteristic of finite-entropy quantum systems.\\

\noindent
Among the recently developed probes of quantum information dynamics, Krylov complexity has emerged as a particularly promising diagnostic. Introduced in \cite{Parker:2018yvk}, Krylov complexity measures the growth of a quantum state or operator in the Krylov basis generated by repeated action of the Hamiltonian through the Lanczos algorithm. Unlike geometric notions of circuit complexity, such as the Nielsen approach \cite{Nielsen:2005mkt} or holographic proposals including complexity=volume \cite{Stanford:2014jda} and complexity=action \cite{Brown:2015bva}, Krylov complexity is intrinsically dynamical and algorithmically well defined. As a result, it has become an increasingly useful tool for studying operator growth, thermalisation, and quantum chaos in both quantum many-body systems and quantum field theory. Indeed, a number of studies have demonstrated that Krylov complexity captures important qualitative distinctions between integrable and chaotic dynamics. In finite-dimensional or integrable systems the complexity typically exhibits bounded or oscillatory behaviour, whereas chaotic systems generically display an initial period of rapid growth followed by late-time saturation at a plateau; see for example \cite{Balasubramanian:2022tpr}. Such behaviour naturally suggests applications to black-hole physics, where one expects an interplay between early-time scrambling and late-time equilibration associated with finite entropy and Hawking radiation. The purpose of this paper is to investigate these questions in a simple quantum mechanical model describing a black hole coupled to its radiation. Our starting point is the proposal of Maldacena \cite{Maldacena:2023acv} that matrix quantum mechanics provides perhaps the simplest nonperturbative framework for describing near-extremal black holes in warped AdS$_2$. More specifically, we consider a simplified ``mini-BMN" matrix model inspired by the BMN deformation \cite{Berenstein:2002jq} of the BFSS matrix model \cite{Banks:1996vh}. Simplified mini-BMN truncations have previously been studied in \cite{Asplund:2015yda,Anous:2017mwr,Han:2019wue,Komatsu:2024vnb,Fliss:2025kzi}, where they were shown to retain a number of the qualitative dynamical features of the full matrix model while remaining considerably more tractable.\\

\noindent
Our goal here is not to reproduce the full dynamics of the BMN matrix model, but rather to isolate a controlled setting in which the information-theoretic properties of a radiating black hole can be studied explicitly. To this end, we consider a reduced matrix quantum mechanics coupled to an additional scalar degree of freedom that plays the role of black-hole radiation. While obviously highly simplified, the resulting system is sufficiently rich to exhibit nontrivial Krylov dynamics and late-time saturation behaviour. The analysis then naturally separates into two complementary regimes.\\

\noindent
At early times, we study the growth of Krylov complexity and Krylov entropy using perturbative and numerical methods. We employ several equivalent formulations of Krylov dynamics, including the Lanczos algorithm itself, spectral moments of two-point functions, and survival amplitudes. These complementary perspectives allow us to characterize the initial growth regime and relate it to expectations from chaotic quantum dynamics.\\

\noindent
At late times, our focus shifts to the emergence of plateau behaviour. Physically, such saturation is to be expected for a black hole in equilibrium with its radiation inside an asymptotically AdS-like setting, where radiation can effectively return from the boundary and re-interact with the black hole. We argue that this late-time regime admits a semiclassical interpretation in terms of Euclidean instanton contributions in an effective path-integral. In particular, building on the framework developed in \cite{Beetar:2025erl}, we show that the plateau value may be understood in terms of nonperturbative saddles of an effective Krylov action. More broadly, the motivation for this work is twofold. First, we wish to understand to what extent Krylov complexity and Krylov entropy provide faithful probes of black-hole information dynamics beyond the regime of early-time scrambling. Second, we aim to develop calculational tools, particularly path-integral and semiclassical methods, that may eventually allow Krylov dynamics to be studied directly in more realistic gravitational systems.\\

\noindent
The paper is organized as follows. In section 2 we construct the simplified matrix model and discuss its relation to the BMN matrix model and its proposed gravitational interpretation. In section 3 we review the necessary aspects of Krylov complexity, Krylov entropy, and related constructions. In section 4 we analyze the early-time dynamics perturbatively and numerically. In section 5 we study the late-time regime, including plateau behaviour, scaling arguments, and the role of Euclidean instantons within an effective path-integral description. We conclude in section 6 with a discussion of the implications of our results and possible future directions.

\section{Quantum mechanical model for a radiating black hole}
We wish to study complexity in a quantum mechanical model that is sufficiently simple to allow for explicit numerical investigation. Krylov complexity is particularly well suited to this purpose in that it is not only algorithmically well defined through the Lanczos construction (though alternative formulations will also prove useful, as reviewed in the next section), but it also naturally gives rise to the associated idea of Krylov entropy. The latter provides an information-theoretic measure whose behaviour may be compared with general expectations for black-hole entropy dynamics, including those suggested by the Page curve.\\

\noindent
It was argued in \cite{Maldacena:2023acv} that the simplest quantum mechanical model capturing essential aspects of black-hole dynamics is the BMN matrix model \cite{Berenstein:2002jq} in an appropriate regime. The BMN model itself is a deformation of the BFSS matrix model \cite{Banks:1996vh}, and may be viewed as a quantum mechanical system of interacting bosonic and fermionic oscillators with action
\bea
S&=& S_B+S_F+S_\omega\cr
S_B&=& \int \sum_{a=1}^ {N^ 2-1}\left[\sum_{I=1}^ 9\frac{1}{2}(\dot X^ {aI})^ 2
-\frac{1}{4}\frac{\lambda}{N}\sum_{I,J=1}
^ 9\left(\sum_{b,c=1}^ {N^ 2-1}{f^ a}_{bc}X^ {Ia}X^ {Jb}\right)^ 2\right]\cr
S_F&=& \frac{N}{\lambda}\int dt \left[\frac{1}{2}\psi^ {\a a}\dot \psi^ {\a a}+\frac{i}{2}\psi^ {\a a}\Gamma^ I_{\a\b}
\psi^{\b b}X^ {I c}{f^ a}_{bc}\right]\cr
&=&\frac{N}{\lambda} \int dt \Tr\left[\frac{1}{2}\psi^\a \dot \psi^ \a +\frac{1}{2}\psi^ \a
\Gamma^ I_{\a\b}[\psi^ \b,X^ I]\right]\cr
S_\omega&=& -\frac{N}{\lambda}\int dt\Tr\left[2\omega^ 2\sum_{I=1}^ 3 (X^ I)^ 2
+\frac{\omega^ 2}{2}\sum_{I=4}^ 9(X^ I)^ 2
+\frac{3i}{4}\omega \psi \Gamma^ 1\Gamma^2 \Gamma^ 3\psi \right.\cr
&&\left.+2i\omega\sum_{I,J,K=1}^ 3
\epsilon_{IJK}X^ I X^ JX^ K\right]\;,\cr
&&
\eea
\noindent
where $\lambda =g^ 2N$, $\{\Gamma^ I,\Gamma_J\}=\delta^ {IJ}$ and the $\Gamma^I$ are 
$SO(9)$ gamma matrices. 
The bosonic term can be rewritten as 
\be 
S_{ B}= \frac{N}{\lambda}\int dt \Tr\left\{\sum_{I=1}^ 9\frac{1}{2}(\dot X^ I)^ 2
+\frac{1}{4}\sum_{I,J=1}^ 9[X^ I,X^ J]^ 2\right\}\,.
\ee
Note that in one spacetime dimension the Yang–Mills coupling $g^2$, and hence the ’t Hooft coupling $\lambda = g^2N$, carries mass dimension three. As a result, the matrix scalars $X^I$ and fermions $\psi^\alpha$ inherit the same canonical scaling dimensions as scalar and fermion fields in four-dimensional field theory. The Hamiltonian of the model is then 
\bea
H&=& \sum_1\left[\frac{1}{2}\sum_{I=1}^ 9 p^ 2_{aI}+\frac{\lambda}{N}\frac{1}{4}\left(\sum_{b,c}\sum_{I,J=1}^ 9
{f^ a}_{bc}x^ {Ic}x^ {Jb}\right)^ 2\right.\cr
&&\left.+\sqrt{\frac{\lambda}{N}}\frac{i}{2}\sum_{\a,\b=1}^{16}\sum_{I=1}^ 9
\sum_{b,c}\psi^ {\a a}
\Gamma^ I_{\a\b}\psi^ {\b b}x^ {Ic}{f^ a}_{bc}+\frac{\omega^ 2}{2}\sum_{I=4}^ 9 x^ 2_{aI}
+\frac{(2\omega)^ 2}{2}\sum_{I=1}^ 3 
x^ 2_{aI}\right.\cr
&&\left.+\frac{3i}{4}\omega
\psi^ {\a a}(\Gamma^ 1\Gamma^2\Gamma^ 3)_{\a\b}\psi^{\b a}-\sqrt{\frac{\lambda}{N}}\omega
\sum_{I,J,K=1}^ 3\sum_{b,c}\epsilon_{IJK}x^ {aI }x^ {bJ }x^ {cK}{f^ a}_{bc}\right].\cr
&&
\eea
\noindent
In order for the model to describe a black hole, we must  consider the large-$N$ and strong-coupling limit $N \gg 1$ and $\omega \ll T \ll \lambda^{1/3}$.
As emphasized in \cite{Maldacena:2023acv}, while we could formally consider the limit $\omega \to 0$, doing so obscures the relevant physics; it is therefore important to work with the BMN deformation rather than the undeformed BFSS model.\\

\noindent
The condition
$T \ll \lambda^{1/3}$
corresponds to the strong-coupling regime, or equivalently
$\lambda/T^3 \gg 1$. In addition, the validity of the supergravity description for D0-branes requires the further condition \cite{Itzhaki:1998dd}
\begin{eqnarray}
    T \gg \lambda^{1/3}N^{-5/9}\,.
    \label{TvsN}
\end{eqnarray}
Together, these inequalities define the regime in which the system admits a near-horizon, near-extremal D0-brane description. In analogy with the finite-temperature AdS/CFT construction of Witten \cite{Witten:1998zw}, the corresponding geometry may be viewed as a black hole in warped $AdS_2 \times S^8$. In this sense, the BMN matrix model provides a quantum mechanical description of a black hole in warped $AdS_2 \times S^8$, and is argued to constitute the simplest model capturing the full dynamics of such a system.\\

\noindent
There are, however, two important subtleties. First, the BMN deformation breaks the $SO(9)$ symmetry of the transverse space down to $SO(3)\times SO(6)$,
so the gravitational dual must be modified accordingly. Second, the full BMN matrix model remains too complicated for the detailed analytic and numerical study of Krylov dynamics that we wish to perform. This motivates the search for a simpler truncation that nevertheless preserves the essential physical features of the original system. As we will argue in section \ref{sec:gravity dual}, there exists a corresponding gravitational regime that captures precisely such a simplification.\\

\noindent
As articulated in the Introduction, our primary interest is not the full microscopic dynamics of the BMN model, but rather the behaviour of Krylov complexity and Krylov entropy as functions of time. In an AdS setting, one expects that at sufficiently late times the black hole and its radiation reach thermal equilibrium, leading to saturation of information-theoretic observables. Indeed, for large black holes in AdS, the temperature increases with mass, leading to a positive specific heat, and the black-hole phase dominates thermodynamically, with $F_{\rm AdS-BH} < F_{\rm AdS}$, thus implying thermodynamical stability, as shown by Witten \cite{Witten:1998zw}. This in turn suggests that the interaction between the black hole and its radiation may be captured by an appropriately extended quantum mechanical model. At the same time, it motivates the construction of a further truncation of the BMN matrix model that is sufficiently tractable for explicit calculations while still retaining the qualitative features relevant for the study of Krylov dynamics.

\subsection{Heat bath and finite temperature}

A simple way to model the heat bath, or equivalently, the Hawking radiation emitted by the black hole, is to couple the matrix quantum mechanics to an additional massless scalar degree of freedom. In the gravitational description, this scalar corresponds to a massless adjoint field in warped $AdS_2 \times S^8$ that becomes thermally excited in the black-hole background. The bosonic kinetic action is then modified to
\be 
S'_{ B}= \frac{N}{\lambda}\int dt \Tr\left\{\sum_{I=1}^ 9\frac{1}{2}(\dot X^ I)^ 2
+\frac{1}{4}\sum_{I,J=1}^ 9[X^ I,X^ J]^ 2
+\frac{(\dot \phi)^ 2}{2}\right\}.\label{kineticdef}
\ee
As we discuss in more detail in section \ref{sec:towards states}, it is in fact necessary to introduce a small mass term for $\phi$, at least as an infrared regulator. This modification should properly be included in $S’_\omega$ in eq. \eqref{BMNdef}. We will also consider the case of finite scalar mass in the perturbative analysis of section \ref{sec:perturbative toy model}.
The scalar $\phi$ must couple to the BMN degrees of freedom in a way that preserves the dimensional structure of the theory. Assigning $\phi$ its canonical four-dimensional scaling dimension, namely $[\phi]=1$, the simplest interaction consistent with dimensional analysis couples $\phi$ to a cubic scalar operator. A convenient implementation is obtained by shifting the BMN deformation parameter according to
$i\omega \;\rightarrow\; i\omega + \tilde g \phi$,
where $\tilde g \sim \mathcal{O}(1)$ is dimensionless. The BMN deformation term is then modified to
\bea
S'_\omega&=& -\frac{N}{\lambda}\int dt\Tr\left[2\omega^ 2\sum_{I=1}^ 3 (X^ I)^ 2
+\frac{\omega^ 2}{2}\sum_{I=4}^ 9(X^ I)^ 2\right.\cr
&&\left.+\frac{3i}{4}\omega \psi \Gamma^ 1\Gamma^2 \Gamma^ 3\psi +(2i\omega+\tilde g\phi)
\sum_{I,J,K=1}^ 3\epsilon_{IJK}X^ I X^ JX^ K\right]\;.\label{BMNdef}
\eea
Finite temperature is introduced in the usual way through the thermal density matrix
\begin{eqnarray}
    \hat\rho = \frac{1}{Z}e^{-\beta \hat H}, \qquad Z = \Tr(e^{-\beta \hat H})\,,
\end{eqnarray}
so that expectation values are replaced according to $\langle \hat A \rangle \;\rightarrow\; \Tr(\hat\rho \hat A)$. Equivalently, and often more conveniently, we may work in the Thermo-Field Double (TFD) formalism, which we will discuss later.

\subsection{Gravity dual with $SO(3)\times SO(6)$ invariance and mini-BMN model}\label{sec:gravity dual}

The BFSS matrix model describes the dynamics of a system of $N$ D0-branes and is conjectured to provide a nonperturbative definition of M-theory in the discrete light-cone quantization (DLCQ) limit. It is important to note, however, that this Matrix-theory limit differs from the holographic gravity-dual limit relevant for the black-hole regime that we will consider below.

\subsubsection{Review of BFSS gravity dual}

The solution for $N$ D0-branes in string frame is given by
\bea
ds_s^2&=&H_0^{-1/2}(-dt^2)+H_0^{1/2}(d\vec{x}_{(9)}^2)\cr
e^{\phi}&=&H_0^{-(0-3)/4}\;,\;\;\; A_0=-\frac{1}{2}(H_0^{-1}-1)\cr
H_0&=&1+d_0\frac{g_{YM}^2N}{U^7\a'^2}\;,\;\;\; d_0=2^7 \pi^{9/2}\Gamma(7/2)\;,
\eea
where $U=r/\a'$, $r^2=\vec{x}_{(9)}^2$, and one makes it non-extremal by introducing a ``blackening" factor
\be
   h(U)=1-\frac{U_0^7}{U^7}\;,
\ee
so that the metric for the $N$ D0-branes at finite temperature is
\be
ds_s^2=H_0^{-1/2}h(-dt^2)+H_0^{1/2}\left(\frac{dr^2}{h}+r^2d\Omega_8^2\right)\;.
\ee
For the gravity dual, we instead consider the near-horizon decoupling limit of  \cite{Itzhaki:1998dd}, 
\be
\a'\rightarrow 0\;,\;\; r\rightarrow 0\;\;\; U={\rm fixed}\;,\;\; g^2_{YM}=\frac{g_s/(4\pi^2)}{\a'^{3/2}}={\rm fixed}\;,
\ee
together with a simultaneous near-extremal limit, 
such that the non-extremality function $h(U)$ is unchanged. The near-horizon near-extremal solution whose field theory dual is the BFSS Lagrangian 
at finite temperature $T$ is then 
\bea
\frac{ds_s^2}{\a'^2}&=&-\frac{U^{7/2}}{\sqrt{d_0}\sqrt{g^2_{YM}N}}h(U)dt^2+\frac{\sqrt{d_0}\sqrt{g^2_{YM}N}}{
U^{7/2}}\left(\frac{dU^2}{h(U)}+U^2d\Omega^2_8\right)\cr
e^{\phi}&=& 4\pi^2g^2_{YM}\left(\frac{d_0g^2_{YM}N}{U^7}\right)^{3/4}\cr
A_0&=& -\frac{1}{2}\left(\frac{\a'^2U^2}{d_0g^2_{YM}N}-1\right).
\eea
Introducing the dimensionless effective coupling
\begin{eqnarray}
    \tilde\lambda = \frac{g^2_{YM}N}{U^3},
\end{eqnarray}
we observe that, if the residual $U$-dependence of $\tilde\lambda$ is temporarily ignored, the geometry takes the form of $AdS_2 \times S^8$ with radius
\begin{eqnarray}
    R^2 = \sqrt{d_0 \tilde\lambda},
\end{eqnarray}
while the dilaton is given by
\begin{eqnarray}
    e^\phi = \frac{4\pi^2}{N}\tilde\lambda^{7/4}\,.
\end{eqnarray}
For physical applications, however, it is more appropriate to work in the Einstein frame rather than the string frame. The two metrics are related by
\be
ds_E^2 = e^{-\phi/2} ds_s^2\,,
\ee
which introduces an additional factor of $U^{21/8}$ into the metric. At large $U$, the resulting Einstein-frame geometry can be rewritten in the form of a warped $AdS_2 \times S^8$ spacetime. This becomes manifest after introducing a new radial coordinate $\rho$ defined through the scaling relations
\bea
e^\phi \sim U^{-21/4}
&\sim&
\rho^{-21/10},
\cr
U^{1/2}
&=&
\rho^{1/5},
\cr
U^{7/2+21/8}
&=&
\rho^2 \rho^{9/20},
\qquad
U^{-3/2+21/8} = \rho^{9/20},
\cr
U^{9/8}\frac{dU^2}{U^2}
&\sim&
\rho^{9/20}\frac{d\rho^2}{\rho^2}.
\eea
In these coordinates, the metric acquires the characteristic form of warped $AdS_2 \times S^8$.
In $\rho$ coordinates, and after a constant rescaling of $t$ to $\tau$, we obtain 
\bea
ds_E^2&=&\tilde \lambda^{-3/8}\frac{\sqrt{d_0N}}{2\pi}
\left[\frac{4}{25}\left(-\rho^2h(\rho)d\tau^2+\frac{d\rho^2}{\rho^2h(\rho)}\right)+d\Omega_8^2\right]\cr
&=&(g^2_{YM}N)^{-3/8}\frac{\sqrt{d_0 N}}{2\pi}\rho^{9/20}
\left[\frac{4}{25}\left(-\rho^2h(\rho)d\tau^2+\frac{d\rho^2}{\rho^2h(\rho)}\right)+d\Omega_8^2\right]\cr
e^\phi&=&\frac{4\pi^2\tilde \lambda^{7/4}}{N}=\frac{4\pi^2(g^2_{YM}N)^{7/4}}{N}\rho^{-21/10}\cr
h&=&1-\frac{U_0^7}{U^7}=1-\left(\frac{\rho_0}{\rho}\right)^{14/5}.
\eea
Note that at $\rho\rightarrow\infty$ the solution becomes warped $AdS_2\times S^8$, so we have 
a D0-brane black hole inside warped $AdS_2\times S^8$. The temperature of the black hole is 
(defined as inverse 
periodicity in $t$, not $\tau$)
\be
\frac{T}{K\lambda^{1/3}}=\frac{[\rho^2 h(\rho)]'}{4\pi}=\frac{7}{5}\frac{\rho_0}{2\pi}\;,
\ee
and the entropy is $\propto$ Horizon Area $\propto [\rho_0^{9/40}]^8$, giving
(here $\tilde K$ is a numerical factor proportional to $N^2$, since $1/G_{N,10}=8\pi M_{\rm Pl,10}^2$ contains 
$N^2$ via AdS/CFT; or rather, that the Einstein metric $ds_E^2$ contains $\sqrt{N}$, so $N^2$ for an 
8-volume)
\be
S=\frac{\rm Area}{4G_{N,10}}=\tilde K \left(\frac{T}{\lambda^{1/3}}\right)^{9/5}.
\ee
The supergravity approximation, needed for the validity of the above gravity dual, 
is valid if ($R^2/\a'\gg 1$, $g_s\ll 1$, so)
\be
g^{2/3}_{YM}N^{1/7}\ll U\ll g^{2/3}_{YM}N^{1/3}=\lambda^{1/3}.
\ee
If we put the temperature scale to be of the order of the energy scale, $T\sim U$, 
the conditions on the temperature are (one from the above, one from (\ref{TvsN}))
\be
T\ll \lambda^{1/3}\;,\;\; T\gg \lambda^{1/3}N^{-5/9}.
\ee
At this point, we should note that the AdS/CFT decoupling limit described above is different from the 
Matrix theory limit for BFSS (DLCQ of M theory). \footnote{There one takes the 11-dimensional radius
$r\rightarrow \infty$, keeping $r/l_P$ fixed instead of $r/\a'$, and takes $N\rightarrow\infty$. In this case the 't Hooft coupling
\be
\lambda_{\rm 't Hooft}=g^2_{\rm YM}N=\frac{Nl_P^3}{r^3}\;,
\ee
is not kept fixed and large as in AdS/CFT, but is instead infinite (or at least much 
larger). Further, in BFSS we are at much lower energies than in AdS/CFT, since 
\be
\frac{r}{l_P}=\frac{U}{g^2_{\rm YM}}\;,
\ee
and $U$ is fixed  because $r/l_P$ is fixed and of order 1, 
as is $g^2_{\rm YM}$ in the BFSS limit, but in AdS/CFT 
we saw above 
that we need instead a much larger $U$, namely
\be
\frac{U}{g^2_{\rm YM}}\gg N^{1/7}.
\ee
Consequently, the gravity dual limit is different from the BFSS limit, as expected.}

\subsubsection{BMN deformation gravity dual and mini-BMN}

Up to this point, the construction is straightforward. The situation changes, however, once the BMN deformation parameter $\omega$ is introduced. In order for the deformation to remain a small perturbation that does not qualitatively alter the thermodynamics, we need $T \gg \omega$. This condition alone is not sufficient. The BMN deformation corresponds to placing the D0-branes in the maximally supersymmetric pp-wave background, itself obtained as an infinite-boost limit of $AdS_4 \times S^7$ in M-theory. As a consequence, the deformation breaks the transverse rotational symmetry according to $SO(9) \rightarrow SO(3)\times SO(6)$. This is incompatible with the original D0-brane geometry based on an $S^8$ transverse space, and therefore the gravitational background must be modified accordingly. Since $SO(9)$ is the isometry group of $S^8$, the simplest geometric realization of this symmetry breaking is to replace
\begin{eqnarray}
    S^8 \rightarrow S^2 \times S^5 \times \tilde r,
\end{eqnarray}
where $\tilde r$ denotes an additional radial direction. In terms of the flat transverse space coordinates, the original decomposition $(r,S^8)$ is thus replaced schematically by two sectors, $(r,S^2)$ and $(\tilde r,S^5)$, with $r$ continuing to play the role of the spatial radial coordinate of the $AdS_2$ factor.\\

\noindent
One may then ask under what conditions the resulting gravitational description can be simplified further so as to capture not the full BMN matrix model, but rather a supersymmetric ``mini-BMN" truncation. Such models were introduced in \cite{Asplund:2015yda,Anous:2017mwr} and subsequently studied numerically in \cite{Han:2019wue,Komatsu:2024vnb,Fliss:2025kzi}, where they were found to retain a number of the qualitative features of the full BMN dynamics while remaining substantially more tractable.
The bosonic mini-BMN model consists of three $SU(N)$ matrix-valued scalars $X^I$. A minimal supersymmetric extension may be constructed by retaining only four supercharges, corresponding to an $SO(3)$ spinor representation. On-shell, the fermionic sector then carries two physical degrees of freedom, matching the two propagating bosonic degrees of freedom obtained from the three scalars after gauge fixing. The bosonic part of the action is
\begin{eqnarray}
    S_B &=& \frac{N}{\lambda}\int dt \, \mathrm{Tr}\Bigg\{
    \sum_{I=1}^3 \frac{1}{2}(\dot X^I)^2 +\frac{1}{4}\sum_{I,J=1}^3 [X^I,X^J]^2
\cr
&&\qquad\qquad
-2\omega^2\sum_{I=1}^3 (X^I)^2
-2i\omega\sum_{I,J,K=1}^3 \epsilon_{IJK}X^I X^J X^K
\Bigg\},
\end{eqnarray}
while the fermionic contribution takes the form
\be
S_F =
\frac{N}{\lambda}\int dt\,, \Tr\left[
\frac{1}{2}\psi^\alpha \dot\psi^\alpha
+\frac{1}{2}\psi^\alpha \Gamma^I_{\alpha\beta}[\psi^\beta,X^I]
-\frac{3i}{4}\omega\psi \Gamma^1\Gamma^2\Gamma^3\psi
\right].
\ee
Here $\Gamma^I$, with $I=1,2,3$, denote the $SO(3)$ gamma matrices, while $\psi^\alpha$ transforms as an $SO(3)$ spinor. Together, these terms define the ${\cal N}=1$ supersymmetric mini-BMN model, which may be viewed as a matrix quantum mechanics associated with an effective 3+1-dimensional target-space structure.\\

\noindent
We are therefore led to a twofold problem. First, we need a gravitational dual whose symmetry is reduced from $SO(9)$ to $SO(3)\times SO(6)$, in accordance with the BMN deformation. Second, we would like this dual description to admit a regime more naturally associated with the mini-BMN truncation than with the full BMN matrix model. The natural way to proceed is to construct the solution in eleven dimensions, as in the standard D0-brane construction. In that case, the D0-brane geometry is obtained by starting with a momentum-carrying Aichelburg–Sexl pp-wave in eleven dimensions and then reducing to ten dimensions. Here we require an analogous construction; the solution should contain a momentum wave, but should also distinguish the effective 3+1-dimensional sector relevant to the mini-BMN model.\\

\noindent
This structure is suggested by the origin of the maximally supersymmetric pp-wave background underlying the BMN deformation. That pp-wave arises as an infinite-boost, or Penrose, limit of $AdS_4\times S^7$, with the boost taken along an equator of $S^7$. To obtain the D0-brane geometry appropriate to the BMN-deformed theory, we should therefore consider a momentum wave propagating along this equatorial direction. Geometrically, this deformation breaks the transverse $S^8$ into
$S^8 \;\longrightarrow\; S^2 \times S^5 \times \tilde r$,
while preserving the $AdS_2$ factor built from the D0-brane time direction and the radial coordinate. Thus the desired construction is an eleven-dimensional pp-wave on $AdS_4\times S^7$, propagating along an equator of $S^7$, followed by reduction to ten dimensions using the standard Kaluza–Klein ansatz. The resulting D0-brane background should then possess precisely the required $SO(3)\times SO(6)$ symmetry.\\

\noindent
From the holographic perspective, this solution should be interpreted as a flow from an ultraviolet geometry of the form $AdS_4\times S^7$ to an infrared geometry of the form
$AdS_2\times S^2\times (\tilde r\times S^5)_{\rm small}$,
where the six-dimensional factor in parentheses becomes small in the infrared. Equivalently, the construction may be viewed as a supersymmetric flow in the dual 2+1-dimensional CFT, plausibly related to the ABJM theory, between two distinct vacua.\\

\noindent
Motivated by the discussion above, we consider the following ansatz for a pp-wave propagating on $AdS_4\times S^7$ along an equatorial direction of $S^7$,
\bea
ds_{11}^2 &=&
R^2\Big[
-\cosh^2 r \, d\tau^2
+dr^2
+\sinh^2 r\, d\Omega_2^2
\cr
&&\qquad
+4\left(
\cos^2\theta\, d\psi^2
+d\theta^2
+\sin^2\theta\, d\Omega_5^2
\right)
\cr
&&\qquad
+(H_0-1)(d\tau+d\psi)^2
\Big],
\cr
F_{t123} &=& \mu,
\label{solution}
\eea
where $(r,\Omega_2)$ parameterize the spatial directions of $AdS_4$, $H_0$ is an as-yet undetermined function, and $\psi$ denotes the equatorial direction along which the pp-wave propagates. The resulting eleven-dimensional geometry may then be reduced along $\psi$ to obtain a ten-dimensional type-IIA background. A fully backreacted black-hole solution with the required $SO(3)\times SO(6)$ symmetry was constructed numerically in \cite{Costa:2014wya}. Their ansatz takes the form
\bea
ds^2 &=&
-A \frac{1-y^7}{y^7}d\tau^2
+T_4 y^7\left[d\psi+\Omega\frac{1-y^7}{y^7}\right]^2
\cr
&&
+\frac{1}{y^2}\Bigg[
B\frac{(dy+Fdx)^2}{(1-y^7)y^2}
+T_1\frac{4dx^2}{2-x^2}
\cr
&&\qquad\qquad
+T_2x^2(2-x^2)d\Omega_2^2
+T_3(1-x^2)^2d\Omega_5^2
\Bigg],
\cr
C_3 &=& (Md\tau+Ld\psi)\wedge d\Omega_2,
\eea
where $x$ is an angular coordinate with $0\le x\le1$ and $y$ is a radial coordinate ranging from the ultraviolet boundary at $y=0$ to the infrared black-hole horizon at $y=1$. In our notation, one may roughly identify $x \sim \theta$,
and $r \sim y^{-1}-1$.
The functions
$A,B,F,\Omega,T_1,T_2,T_3,T_4,M,L$
depend on both $x$ and $y$. The ultraviolet boundary conditions were chosen such that
\begin{eqnarray}
    A,B,\Omega,T_1,T_2,T_3,T_4 = 1+\mathcal O(y), \qquad F=\mathcal O(y),
\end{eqnarray}
while $M$ and $L$ scale respectively as $y^{-3}$ and $y^4$ times fixed functions of $x$. For our purposes, however, the full numerical solution is not required. We will only need its behaviour near the infrared horizon region, where our simplified ansatz agrees with the general solution, together with certain features of the ultraviolet asymptotics that we discuss later.\\

\noindent
In Appendix A we show that the ansatz \eqref{solution} indeed solves the equations of motion provided that $H_0$ satisfies a Poisson equation on $AdS_4\times S^7$. In the infrared region, where the geometry locally reduces to flat space, this equation becomes
\be
\Delta_{\vec{z}_{(9)}=(r,\Omega_2,\theta,\Omega_5)}H_0=Q\delta^9(\vec{z}_{(9)})\;,\label{harmonic}
\ee
This approximation is valid near the centre of $AdS_4$ and close to an equator of $S^7$, namely near $r=0$, and $\theta=0$,
which is precisely the region in which the maximally supersymmetric BMN pp-wave emerges. In this limit, the metric reduces approximately to
\be
ds_{11}^2 \simeq
R^2\Big[
-d\tau^2
+dr^2
+r^2 d\Omega_2^2
+4(d\psi^2+d\theta^2+\theta^2 d\Omega_5^2)
+(H_0-1)(d\tau+d\psi)^2
\Big].
\label{approxsol}
\ee
It is useful to note that, had we instead set $Q=0$, retained the next-order corrections in $r$ and $\theta$, and taken the Penrose limit around the same null geodesic, we would recover the standard maximally supersymmetric pp-wave background. Here, however, we proceed differently by solving the Poisson equation \eqref{harmonic} in the locally flat geometry above, obtaining
\be
H_0 \simeq
1+\frac{C_{11}Q/2}{(r^2+\theta^2)^{7/2}}.
\ee
Substituting this back into \eqref{approxsol} yields the eleven-dimensional Aichelburg–Sexl pp-wave geometry
\be
ds_{11}^2
=
2dx^+dx^-
+(H_0-1)(dx^+)^2
+d\vec x_{(9)}^2,
\label{11dwave}
\ee
where
\begin{eqnarray}
    x^\pm = \frac{\psi\pm\tau}{\sqrt2}, \qquad \tilde r^2 = r^2+\theta^2.
\end{eqnarray}
Finally, reducing along the $\psi$ direction using the standard Kaluza–Klein ansatz
\be
ds_{11}^2
=
e^{-2\phi/3}ds_{s,10}^2
+
e^{4\phi/3}(dx^{11}+A_\mu dx^\mu)^2,
\ee
with $x^{11}=\psi$, produces the ten-dimensional type-IIA background corresponding to a system of $N$ D0-branes in flat space.\\

\noindent
Since we are interested in finite-temperature configurations, the next step is to introduce a non-extremal deformation of the geometry through an appropriate blackening factor. Crucially, this deformation must preserve the reduced $SO(3)\times SO(6)$
symmetry implied by the BMN background. Geometrically, this amounts to breaking the $SO(3,2)$ isometry of $AdS_4$ down to the
$SO(1,2)\times SO(3)$ symmetry of $AdS_2\times S^2$. To achieve this, we introduce a Schwarzschild-like blackening factor depending only on the radial coordinate $r$,
\be
  h(r)=1-\frac{2MG_N}{r},
\ee
and insert it only into the $(t,r)$ part of the metric, leaving the transverse $\theta$ directions unaffected. The resulting ten-dimensional string-frame geometry becomes
\be
   ds_{s,10}^2=
   -H_0(\tilde r)^{-1/2}h(r)dt^2
   +
   H_0(\tilde r)^{1/2} \left( \frac{dr^2}{h(r)} +r^2d\Omega_2^2 +d\theta^2 +\theta^2d\Omega_5^2 \right)\,,
\ee
with $F_{t123}=\mu$ as before. This differs from the standard black D0-brane solution, for which $\mu=0$ and the blackening factor depends on the full transverse radius $\tilde r$ rather than only on $r$. Passing to the Einstein frame,
\be
   ds_{E,10}^2 = H_0^{-3/8}(\tilde r)\,ds_{s,10}^2\,,
\ee
we now take the usual near-horizon decoupling limit
$r\rightarrow0$, $\theta\rightarrow0$,
for which
\be
   H_0 \simeq \frac{C_{11}Q/2}{(r^2+\theta^2)^{7/2}}\,.
\ee
At this stage, however, we impose an additional restriction. Since we ultimately wish to perform an effective Kaluza–Klein reduction on the extra dimensions, we focus on the regime $\theta \ll r$.
In this limit the harmonic function simplifies to
\be
   H_0(r) \simeq \frac{C_{11}Q/2}{r^7}\,,
\ee
and the Einstein-frame metric may be written as
\be
   ds_{E,10}^2 = r^2H_0^{1/8}(r) \left[-\frac{H_0^{-1}(r)}{r^2}h(r)dt^2 +\frac{dr^2}{r^2h(r)} +d\Omega_2^2 +\frac{d\theta^2+\theta^2d\Omega_5^2}{r^2}\right].
\ee
As in the standard black D0-brane analysis, we now introduce a new radial coordinate $\rho$ by rewriting the factor
\begin{eqnarray}
    \frac{H_0^{-1}(r)}{r^2} \sim r^{7-2}
\end{eqnarray}
in the form $\sim \rho^2$. In terms of $\rho$, the metric becomes
\be
ds_{E,10}^2
\sim
\rho^{9/20}
\left[
\frac{4}{25}
\left(
-\rho^2 h(\rho)dt^2
+\frac{d\rho^2}{\rho^2 h(\rho)}
\right)
+d\Omega_2^2
+\frac{d\theta^2+\theta^2d\Omega_5^2}{\rho^{4/5}}
\right].
\ee
Taking the limit $\rho\to\infty$, the extra-dimensional sector shrinks away and the geometry approaches warped
$AdS_2\times S^2$.
Thus the infrared theory describes an effectively four-dimensional black hole embedded in warped $AdS_2\times S^2$. In this limit the four-form flux
$F_{(4)}=\mu \epsilon_4$
acts simply as an effective cosmological constant in four dimensions.\\

\noindent
This is precisely the structure expected for the mini-BMN truncation in the sense that the extra dimensions decouple, leaving only the time direction together with the three matrix coordinates $X_1,X_2,X_3$,
which become the effective spatial directions of the reduced matrix quantum mechanics. The resulting symmetry structure is therefore $SO(3)\times SO(6)$,
where the $SO(6)$ factor originates from the compact $S^5$, even though the latter shrinks in the infrared limit.
The thermodynamics also differs from that of the ordinary D0-brane geometry. While the temperature still scales as
$T\propto \lambda^{1/3}\rho_0$,
the entropy acquires a different scaling behaviour because the horizon area is now proportional to $[\rho_0^{9/40}]^2\,
{\rm Vol}_{S^2} \times {\rm Vol}_{6d}$. One therefore finds
\be
S
=
\frac{\rm Area}{4G_{N,10}}
=
\tilde K’
\left(
\frac{T}{\lambda^{1/3}}
\right)^{9/20}
\frac{{\rm Vol}_{6d}}{G_{N,10}}
=
\frac{\tilde K’}{G_{N,4}}
\left(
\frac{T}{\lambda^{1/3}}
\right)^{9/20}.
\label{BMNent}
\ee
It is important to justify the restriction $\theta\ll r$ used above. The key point is that this truncation is consistent within the infrared region of interest. In this regime one may first restrict to small $\theta$, then introduce the non-extremal deformation only in the $r$ direction, and finally perform an effective reduction on the transverse $\theta$ sector. Of course, such a reduction cannot be globally valid for the full geometry, since $\theta$ eventually becomes large. Nevertheless, within the infrared near-horizon region relevant for our purposes, the truncation is self-consistent, and it is precisely this region that reproduces the mini-BMN structure.\\

\noindent
For consistency, one must also examine the ultraviolet behaviour of the solution. As expected, the geometry asymptotes to $AdS_4\times S^7$. In the strict ultraviolet limit the $S^7$ shrinks to zero size, leaving effectively an $AdS_4$ gravitational dual. Away from the exact UV limit, however, the ansatz \eqref{solution} remains valid, and the function $H_0$ continues to satisfy the harmonic equation \eqref{harmonic}, now interpreted as a pp-wave perturbation of the $AdS_4\times S^7$ geometry. As usual, the corresponding solution vanishes asymptotically in the UV. Upon reduction from eleven to ten dimensions along $\psi$, the $S^7$ reduces to the familiar $\mathbb{CP}^3$ geometry.\\

\noindent
In summary, the gravitational dual of the (mini-)BMN model interpolates between a warped $AdS_2\times S^2$ black-hole geometry in the infrared and an asymptotic $AdS_4$ geometry in the ultraviolet, accompanied by an $S^7$, or equivalently a $\mathbb{CP}^3$ after reduction, whose size vanishes asymptotically.

\subsection{BMN Matrix toy model}
\label{sec:BMNtoymodel}

As it stands, the quantum mechanical system defined by the bosonic actions \eqref{kineticdef} and \eqref{BMNdef} remains too complicated for a detailed analytic and numerical study of Krylov dynamics. We therefore introduce a further simplified toy model that, while more tractable, is expected to retain the essential qualitative features relevant for the behaviour of Krylov complexity and Krylov entropy in a black-hole setting.\\

\noindent
Our first simplification is to retain only the three matrix scalars $X^I$, $I=1,2,3$.
In the supersymmetric theory this reduction corresponds precisely to the mini-BMN matrix model discussed above, originally introduced in \cite{Asplund:2015yda,Anous:2017mwr} and studied numerically for the $SU(2)$ case in \cite{Han:2019wue,Komatsu:2024vnb,Fliss:2025kzi}. Our second simplification is to discard the fermionic sector entirely, leaving a purely bosonic model. Since our primary interest lies in the structure of the bosonic background and its associated information-theoretic dynamics, we expect this truncation to preserve the qualitative features most relevant for the study of Krylov complexity and entropy.\\

\Comment{ Then we will consider only $N=3$, which we will take to be enough for 
a "large $N$" expansion, based on the analogous statement in lattice QCD, justified by the fact that 
we have $1/N^2=1/9\simeq 0.1$ corrections, which are small. This will then give $(N^2-1)\times 3=24$ bosonic
oscillators, with which we can do some numerical work.\footnote{Note, however, that in  
\cite{Maldacena:2023acv} it was said that, experimentally, $N=16$
with $16 N^2$ bosonic oscillators and comparable fermionic oscillators gives a reasonable description of a 
black hole.}}

\noindent
Finally, and more drastically, we drop the quartic commutator term, so that the interaction becomes only 
the ($(2i\omega+g\phi)\sum_{I,J,K=1}^ 3\epsilon_{IJK}X^ I X^ J X^ K$) term. This needs some explanation. 
But first, let us present the resulting model via some relevant rescalings, and in the regime needed 
to describe the D0-brane black hole in $AdS_2\times S^8$. The action of our bosonic mini-BMN Matrix model without the quartic interaction is 
\be
S=\frac{N}{\lambda}\int dt\;, \Tr \left\{\sum_{I=1}^3\left[\frac{1}{2}(\dot X^I)^2
-\frac{(2\omega)^2}{2}(X^I)^2\right]+\frac{\dot \phi^2}{2}
-i(2\omega +\tilde g \phi)\sum_{I,J,K=1}^3\epsilon_{IJK}X^I X^J X^K\right\}\;,
\ee
with corresponding Hamiltonian 
\be
H=\frac{N}{\lambda}\Tr\left\{\sum_{I=1}^3\left[\frac{1}{2}(\dot X^I)^2
+\frac{(2\omega)^2}{2}(X^I)^2\right]+\frac{\dot \phi^2}{2}
+i(2\omega+\tilde g\phi)\sum_{I,J,K=1}^3\epsilon_{IJK}X^I X^J X^K\right\}.\label{toyH}
\ee
Our toy model Hamiltonian above, (\ref{toyH}), can be rewritten by rescaling in the schematic form
\bea
H&=&N\lambda^{1/3}\Tr\left\{\sum_{I=1}^3 \left[\left(\frac{dX^I/\lambda^{1/3}}{dt\cdot \lambda^{1/3}}\right)^2
+\left(\frac{\omega}{\lambda^{1/3}}\right)^2
\left(\frac{X^I}{\lambda^{1/3}}\right)^2\right]\right.\cr
&&\left.+\left(\tilde g\frac{\phi}{\lambda^{1/3}}
+\frac{\omega}{\lambda^{1/3}}\right)\left(\frac{X}{\lambda
^{1/3}}\right)^3+\left(\frac{d\phi/\lambda^{1/3}}{dt\cdot \lambda^{1/3}}\right)^2\right\}\;,\label{schemH}
\eea
where all the terms are in terms of dimensionless variables (with the dimension 
defined by $\lambda^{1/3}$, and only the 
overall $\lambda^{1/3}$ has the dimension 1 of the Hamiltonian. 
The conditions defining the BMN D0-brane regime,
\be
\omega \ll T \ll \lambda^{1/3},
\qquad
\lambda^{1/3}N^{-5/9}\ll T,
\label{BMNcond}
\ee
may equivalently be written in dimensionless form as
\be
\frac{\omega}{T}\ll1,
\qquad
\frac{T}{\lambda^{1/3}}\ll1,
\qquad
N\left(\frac{T}{\lambda^{1/3}}\right)^{9/5}\gg1.
\ee
In this regime, the Hamiltonian scales schematically as
\be
H
\propto
N^2\lambda^{1/3}
f\left(
\frac{\omega}{\lambda^{1/3}},
\tilde g
\right),
\ee
so that the corresponding Boltzmann factor behaves as
\be
e^{-\beta H}
\propto
\exp\left[
N^2
\left(
\frac{\lambda^{1/3}}{T}
\right)
f\left(
\frac{\omega}{\lambda^{1/3}},
\tilde g
\right)
\right].
\ee
Since both
\begin{eqnarray}
    \frac{\lambda^{1/3}}{T}\gg1
\qquad\text{and}\qquad
N\gg1,
\end{eqnarray}
the overall coefficient appearing in the exponential is parametrically large. As a consequence, the kinetic contributions, such as $\dot X^2$ and $\dot\phi^2$, dominate the dynamics. By contrast, the quadratic BMN deformation terms proportional to $\omega^2$ are parametrically suppressed, since they enter with an effective coefficient
\be
\frac{\lambda^{1/3}}{T}
\frac{\omega^2}{\lambda^{2/3}}
=
\frac{\omega}{\lambda^{1/3}}
\frac{\omega}{T}
\ll1.
\ee
Similarly, the cubic self-interaction terms proportional to $\omega X^3$ are also suppressed, appearing with coefficient
\be
\frac{\lambda^{1/3}}{T}
\frac{\omega}{\lambda^{1/3}}
=
\frac{\omega}{T}
\ll1.
\ee
Thus, within the regime of interest, both the quadratic BMN mass terms and the intrinsic cubic BMN self-interactions may consistently be treated as subleading contributions. On the other hand, the terms with $\tilde g$ come with coefficient
\be
\tilde g \frac{\lambda^{1/3}}{T}\gg 1\;,
\ee
so they correspond to {\em strong} interactions! \\

\noindent
We now return to the question of whether it is consistent to neglect the quartic commutator interaction in the Hamiltonian, namely the term of the form $[X,X]^2$. Applying the same rescalings as in \eqref{schemH}, this term contributes to $\beta H$ as
\be
\beta H
=
\cdots
+
\frac{N\lambda^{1/3}}{T}
\int
\left[
\frac{X}{\lambda^{1/3}},
\frac{X}{\lambda^{1/3}}
\right]^2.
\ee
Thus, after factoring out the common prefactor, the quartic term is naturally of order one. This seems problematic, since the $\phi X^3$ interaction that we retained is of order $\tilde g$. If $\tilde g\ll1$, then the quartic commutator term appears parametrically larger than the interaction we wish to keep.
This tension is closely related to the fact that the regime of interest is a strongly coupled Yang–Mills regime, dual to a classical gravitational description. As emphasized in \cite{Maldacena:2023acv}, the relevant dimensionless effective coupling at temperature $T$ is
$\frac{\lambda}{T^3}$,
while at zero temperature it is instead controlled by
$\frac{\lambda}{\omega^3}$.
The hierarchy of scales is therefore
\be
1\ll
\frac{\lambda^{1/3}}{T}
\ll
\frac{\lambda^{1/3}}{\omega}.
\ee
We may still choose the radiation coupling to satisfy
\be
\tilde g \gg \frac{\omega}{\lambda^{1/3}},
\ee
so that the intrinsic BMN cubic term proportional to $\omega X^3$ can be neglected, while the radiation-induced interaction $\phi X^3$ is retained. However, this does not by itself justify dropping the quartic commutator interaction, which remains of order one in the strongly coupled regime.\\

\noindent
A possible way to make the truncation consistent is to restrict attention to a low-energy region of the matrix configuration space. In the absence of both the BMN deformation and finite temperature, the effective coupling at an energy scale $E_0$ is
\be
\lambda_{\rm eff}
=
\frac{g^2_{YM}N}{E_0^3}
=
\frac{\lambda}{E_0^3}.
\ee
Suppose we restrict the integration over matrix configurations to the regime
\be
X \leq E_0 \ll \lambda^{1/3},
\ee
so that
$\lambda^{1/3}/E_0\gg1$.
Then the quartic interaction can be made parametrically smaller than the cubic radiation coupling, provided
\be
\tilde g,\frac{\phi}{\lambda^{1/3}}
\gg
\frac{E_0}{\lambda^{1/3}}
\geq
\frac{X}{\lambda^{1/3}}.
\ee
Equivalently, the required condition is
\be
\tilde g \phi \gtrsim E_0.
\ee
In this regime, the $X$ degrees of freedom describing the black-hole sector are restricted to low energies, while $\phi$ represents the radiation sector. By the UV/IR correspondence, this low-energy restriction on the matrix degrees of freedom corresponds on the gravitational side to focusing on a high-energy, or near-boundary, Wilsonian regime. In this sense, the truncated model may be interpreted as an effective description in which the quantum integration is performed only over modes with energies $E\geq E_0$. The remaining question is whether this restricted regime is sufficient for the Krylov-complexity and Krylov-entropy calculations of interest. This is the criterion that the toy model must ultimately satisfy.

\section{Krylov complexity review}\label{sec:krylov review}

As discussed above, our primary focus will be on Krylov complexity and the associated notion of Krylov entropy, both of which can be computed efficiently, at least perturbatively, within the quantum mechanical toy model. Krylov complexity is defined through the decomposition of a time-dependent quantum state into a Krylov basis generated by repeated action of the Hamiltonian, 
\be
|\psi(t)\rangle=\sum_n \psi_n(t)|K_n\rangle\;,
\ee
where the Krylov basis $|K_n\rangle$ is determined by recursively applying the Gram-Schmidt orthonormalization 
procedure to the elements 
\be
|\psi_n\rangle=\hat H^n|\psi(t=0)\rangle\;,
\ee
and 
\bea
|K_n\rangle&=&b_n^{-1}|A_n\rangle\cr
|A_{n+1}\rangle&=& (\hat H-a_n)|K_n\rangle-b_n |K_{n-1}\rangle\;,
\eea
where the Lanczos coefficients $a_n$ and $b_n$ are defined by 
\be
a_n=\langle K_n|\hat H|K_n\rangle\;,\;\; b_n=\sqrt{\langle A_n|A_n\rangle}.
\ee
The initial conditions to the recursive definition are 
\be
|K_0\rangle=|\psi(t=0)\rangle\;,\;\; b_0=0.
\ee
In the Krylov basis the Hamiltonian is tri-diagonal, since the Lanczos algorithm implies 
\be
\hat H|K_n\rangle=a_n|K_n\rangle +b_{n+1}|K_{n+1}\rangle+b_n|K_{n-1}\rangle.
\ee
For the definition of the Krylov complexity, consider a reference state $|\psi(t=0)\rangle \equiv |\psi_0\rangle$, 
and evolve it in time with the Hamiltonian, to get $|\psi(t)\rangle$. Then the Krylov complexity is defined as
\be
K(t)=\langle \psi(t)|\left(\sum_n n|K_n\rangle\langle K_n|\right)|\psi(t)\rangle=
\langle\psi(t)|\hat K|\psi(t)\rangle=\sum_n n |\psi_n(t)|^2\;,\label{Kpsin}
\ee
where we have defined the Krylov complexity operator 
\be
\hat K=\sum_n n|K_n\rangle\langle K_n|.
\ee
A more general notion is the spread complexity, where we replace $n$ with general coefficients $c_n$, so that 
\be
C(t)=\langle \psi(t)|\left(\sum_n c_n |B_n\rangle\langle B_n|\right)|\psi(t)\rangle=\langle \psi(t)|\hat B|\psi(t)\rangle
=\sum_n c_n |\psi_n(t)|^2,
\ee
where (the notation follows  \cite{Haque:2022ncl}, \cite{Chattopadhyay:2023fob})
\be
\hat B=\sum_n c_n|B_n\rangle\langle B_n|.
\ee
Another useful notion is that of the {\em survival amplitude},
\be
S(t)\equiv \langle \psi(t)|\psi(0)\rangle=\langle \psi(0)|e^{iHt}|\psi(0)\rangle\equiv \langle K_0|e^{iHt}|K_0\rangle\;,
\ee
since its moments are related to the Krylov basis $|K_n\rangle$, because
\be
\mu_n=\left.\frac{d^n}{dt^n}S(t)\right|_{t=0}=\langle K_0|(\hat H)^n|K_0\rangle\,,
\ee
recalling that $|K_0\rangle
\equiv|\psi(t=0)\rangle\equiv |\psi_0\rangle$.\\

\noindent
A useful perspective is to associate quantum states to operators. Given a reference state $|\psi\rangle$, one defines the state corresponding to an operator ${\cal O}$ by
\be
|{\cal O}) \equiv {\cal O}|\psi\rangle\,.
\ee
In this way, questions about operator growth may be reformulated as questions about the evolution of states in an associated Krylov space. The natural inner product on this space is
\be
({\cal O}_n|{\cal O}_m)
=
\langle\psi|
{\cal O}_n^\dagger {\cal O}_m
|\psi\rangle,
\ee
and the Krylov basis of operators, denoted $\hat{\cal O}_n$, is generated by applying the Lanczos algorithm to the states $|{\cal O}_n)$. The time-dependent operator $\hat{\cal O}(t)$ may then be decomposed in the Krylov basis as
\be
\hat{\cal O}(t)
\equiv
\sum_n i^n \psi_n(t)\hat{\cal O}_n,
\qquad
\psi_n(t)
=
i^{-n}
(\hat{\cal O}_n|\hat{\cal O}(t)).
\ee
Equivalently, at each time $t$ one may associate to the operator a corresponding state in Krylov space,
\be
|\hat{\cal O}(t))
\equiv
\hat{\cal O}(t)|\psi\rangle.
\ee
At finite temperature, the inner product is modified by replacing vacuum expectation values with thermal expectation values defined using the density matrix
$
\hat\rho
=
\frac{1}{Z}e^{-\beta \hat H}$.
The thermal inner product is therefore
\be
({\cal O}_n|{\cal O}m)\beta
\equiv
\frac{1}{Z}
\Tr\left[
e^{-\beta\hat H}
{\cal O}_n^\dagger
{\cal O}_m
\right]\,.
\ee
The corresponding Krylov coefficients are then obtained from
\be
\psi_n(t)
=
i^{-n}
(\hat{\cal O}n|\hat{\cal O}(t))\beta.
\ee
A complementary way to characterize the Krylov decomposition at finite temperature is through thermal two-point functions. Here, one defines
\be
C(t,\beta)
\equiv
({\cal O}|{\cal O}(-t))_\beta
\equiv
\sum_n M_n \frac{(-it)^n}{n!},
\ee
where the coefficients $M_n$ are the moments of the correlator,
\be
M_n
=
\frac{1}{(-i)^n}
\left.
\frac{d^n C(t,\beta)}{dt^n}
\right|_{t=0}.
\ee
Equivalently, the moments may be expressed in terms of the spectral function $f(\omega)$,
\be
M_n
=
\int_{-\infty}^{+\infty}
\frac{d\omega}{2\pi}
\,\omega^n f(\omega)\,,
\ee
where $f(\omega)$ is the Fourier transform of the thermal correlator,
\be
f(\omega)
=
\int_{-\infty}^{+\infty}
dt\,
e^{i\omega t}
C(t,\beta).
\ee
These moments determine the Lanczos coefficients and hence the corresponding Krylov basis.\\

\noindent
A third and particularly useful formulation at finite temperature is obtained by purifying the thermal density matrix through the Thermo-Field Double (TFD) construction. The thermal mixed state at temperature
$T=\frac{1}{k_B\beta}$
is replaced by the pure TFD state
\be
|\psi\rangle
\rightarrow
|0,\beta\rangle
\equiv
\frac{
\sum_{n=\tilde n}
e^{-\beta E_n/2}
|n\rangle\otimes |\tilde n\rangle
}{
\left(
\sum_m e^{-\beta E_m}
\right)^{1/2}
}\,.
\ee
Following \cite{Balasubramanian:2022tpr}, Krylov complexity may then be studied by evolving the TFD state with the Hamiltonian $\hat H$. Denoting the initial thermal state by
$|0,\beta\rangle \equiv |\psi_\beta\rangle$,
its time evolution is
\be
|\psi_\beta(t)\rangle
=
e^{-iHt}|\psi_\beta\rangle
=
|\psi_{\beta+2it}\rangle.
\ee
The corresponding analytically continued partition function,
\be
Z_{\beta-it}
=
\sum_n e^{-(\beta-it)E_n},
\ee
defines the spectral form factor (SFF),
\be
{\rm SFF}{\beta-it}
\equiv
\frac{|Z{\beta-it}|^2}{|Z_\beta|^2}.
\ee
The survival amplitude of the time-evolved TFD state is directly related to the spectral form factor,
\be
S_\beta(t)
=
\langle\psi_{\beta+2it}|\psi_\beta\rangle
=
\frac{Z_{\beta-it}}{Z_\beta},
\ee
so that
\be
{\rm SFF}_{\beta-it}
=
|S_\beta(t)|^2.
\ee
The moments extracted from the spectral form factor are in turn related to the Krylov basis and therefore encode information about the associated Krylov complexity.\\

\noindent
While Krylov complexity and spread complexity provide useful diagnostics of operator growth, the quantity most directly relevant for black-hole information dynamics is arguably the associated \textit{Krylov entropy}, defined by
\be
S_K(t)
\equiv
-\sum_n |\psi_n(t)|^2
\log |\psi_n(t)|^2.
\ee
Like Krylov complexity itself, the Krylov entropy is constructed from the Krylov-basis coefficients $\psi_n(t)$, and in principle should also be evaluated at finite temperature.
For a pure state
$|\psi\rangle
=
\sum_n \psi_n |K_n\rangle$,
the Krylov entropy takes the form
\be
S_K
=
-\sum_n |\psi_n|^2 \ln |\psi_n|^2.
\ee
By contrast, the usual von Neumann entropy,
\be
S_{\rm vN}
=
-\Tr(\hat\rho\ln\hat\rho),
\ee
vanishes identically for a pure state. Nevertheless, the Krylov entropy may be related to an ordinary entanglement entropy by employing a doubling construction analogous to that used in the Thermo-Field Double formalism. One introduces a doubled Hilbert space ${\cal H}_A\otimes{\cal H}_B$ and maps the Krylov basis states according to
\be
|K_n\rangle
\rightarrow
|K_n\rangle_A
\otimes
|\tilde K_n\rangle_B.
\ee
Tracing over the auxiliary copy $B$ then produces the reduced density matrix
\bea
\rho_A
&\equiv&
\Tr_B \rho
\cr
&=&
\Tr_B\Bigg[
\sum_n \psi_n
|K_n\rangle_A
\otimes
|\tilde K_n\rangle_B
\sum_m
{}_B\langle \tilde K_m|
\otimes
{}_A\langle K_m|
\psi_m^*
\Bigg]
\cr
&=&
\sum_n
|\psi_n|^2
|K_n\rangle\langle K_n|.
\eea
The associated von Neumann entropy is therefore
\be
S_{\rm vN,A}
=
-\Tr(\rho_A\ln\rho_A),
\ee
which coincides exactly with the Krylov entropy,
\be
S_K
=
S_{\rm vN,A}.
\ee
Thus Krylov entropy may be interpreted as an entanglement entropy in the doubled Krylov-space construction. In this sense, it provides a genuine entropy measure associated with the spread of amplitudes across Krylov space.\\

\noindent
From the perspective of black-hole physics, this quantity is particularly interesting. Since Krylov entropy measures the effective number of populated Krylov states, one expects it to increase during the early stages of operator growth, much as Krylov complexity itself typically grows linearly at early times. For a black hole in AdS space, the initial behaviour should closely resemble that of a black hole in asymptotically flat space, since the AdS boundary is not yet dynamically relevant.
At sufficiently late times, however, the situation changes qualitatively. Radiation emitted by the black hole can return from the AdS boundary and fall back through the horizon, eventually establishing thermal equilibrium between the black hole and its radiation bath. Consequently, one expects the total entropy to cease growing and instead approach a constant late-time value. In Krylov-space language, this corresponds to the emergence of a plateau in the Krylov entropy.

\subsection{Examples for motion on group spaces}

For dynamics on a continuous group manifold, one generally expects the complexity not to saturate, but instead to exhibit quasi-periodic or fully periodic behaviour reflecting the underlying integrable structure of the system. This expectation is borne out in the three simple examples discussed below, taken from \cite{Balasubramanian:2022tpr},  for Hamiltonians of the form
\be
H=\a(L_++L_-)+\gamma L_0+\delta \one\,.
\ee
We will begin by computing the Krylov complexity in each case.
\begin{enumerate}
    \item {\em Particle moving in the $SL(2,\mathbb{R})$ group},  with algebra
    \be
      [L_0,L_\pm]=\mp L_\pm\;,\;\; [L_+,L_-]=2L_0
    \ee
    and representation defined by the scaling dimension $h$, as $|h,n\rangle$,
    \bea
      L_0|h,n\rangle&=& (h+n)|h,n\rangle\cr
      L_+|h,n\rangle&=&\sqrt{(n+1)(2h+n)}|h,n+1\rangle\cr
      L_-|h,n\rangle&=& \sqrt{n(2h+n-1)}|h,n-1\rangle.
    \eea
    For evolution starting from the highest weight state, $|h,n\rangle=|K_n\rangle$. For the {\bf harmonic oscillator}, with 
    \be
      Z(\b)= \frac{e^{\b\hbar\omega}}{e^{\b\hbar\omega}-1}\Rightarrow S_\b(t)=\frac{1-e^{-\b\hbar\omega}}{1-e^{-(\b-it)\hbar \omega}}\;,
    \ee
    one finds matching to the above for 
    \be
      h=\frac{1}{2}\;,\;\;\; \gamma=\frac{\b}{\tanh (\b\omega/2)}\;,\;\;
      \delta=-\frac{\omega}{2}\;,\;\;\;
      \a=\frac{\omega}{2\sinh(\b\omega/2)}.\label{harmoscmatch}
    \ee
    Then the initial TFD state is 
    \be  |\psi_\b\rangle=|K_0\rangle=|h=1/2\rangle\;,
    \ee
    and one finds the complexity at temperature $T$ is {\em periodic in time}, 
    \be
      K(t)=\frac{\sin^2(\omega t/2)}{\sinh^2(\b\omega/2)}.
    \ee
    \item For a {\em particle moving in the $SU(2)$ group}, with algebra
    \be
      [L_0,L_\pm]=\pm L_\pm\;,\;\; [L_+,L_-]=2L_0\;,
    \ee
    and Krylov basis for spin $j$, $|K_n\rangle=|j,-j+n\rangle$, so (at zero temperature) 
    $|K_0\rangle=|j,-j\rangle$, time dependence
    \be
      |\psi(t)\rangle=e^{-iHt}|j,-j\rangle\;,
    \ee
    one finds the complexity at $T=0$ that is again {\em periodic in time},
    \be
      K(t)=\frac{2j}{1+\frac{\gamma^2}{4\a^2}}\sin^2\left(\a t\sqrt{1+\frac{\gamma^2}{4\a^2}}\right).
    \ee
    \item {\em Particle moving in the Heisenberg-Weyl group}, 
    of $(L_+=a^\dagger, L_-=a,L_0=N=a^\dagger a)$, with $\a\rightarrow\lambda$, 
    $\gamma\rightarrow \omega$, so 
    \be
      H=\lambda(a^\dagger +a)+\omega N+\delta \one\;,
    \ee
    and initial state $|\psi\rangle=|K_0\rangle=|0\rangle$, so 
    \be
      |\psi(t)\rangle=e^{-iHt}|0\rangle\;,
    \ee
    one finds the complexity at zero temperature that is yet again {\em periodic in time},
    \be
      K(t)=\frac{4\lambda^2} {\omega^2}\sin^2\left(\frac{\omega t}{2}\right).
    \ee
\end{enumerate}
For the Krylov entropy, it will suffice to consider the example of the harmonic oscillator, coming from the motion 
on $Sl(2,\mathbb{R})$. Calculating the Krylov entropy as
\be
S_K=-\sum_n |\psi_n(t)|^2\ln |\psi_n(t)|^2\;,
\ee
we find 
\bea
S_K&=& -\frac{x\ln x+(1-x)\ln (1-x)}{x}\cr
x&=& \frac{1-\frac{\gamma^2}{4\a^2}}{\cosh^2\left(\a t\sqrt{1-\frac{\gamma^2}
{4\a^2}}\right)-\frac{\gamma^2}{4\a^2}}.
\eea
Then, for $\gamma\neq 2\a$, we find {\em at late times} $t$
\be
S_k\simeq 2\a t\sqrt{1-\frac{\gamma^2}{4\a^2}}\;,
\ee
whereas if $\gamma=2\a$ {\em exactly} (as is the case for the identification to the 
harmonic oscillator {\em at $\b\omega
\rightarrow 0$}), we have 
\be
S_K\sim 2\ln (\a t)+1+...
\ee
Note, however, that in our case
$\beta\omega=\frac{\omega}{T}\ll1$,
though not exactly zero. More importantly, for the harmonic oscillator one generally finds
$\frac{\gamma}{2\alpha}\geq1$,
so that the formal expression obtained above becomes imaginary. Strictly speaking, this means that the result cannot be interpreted literally within the present framework. One possible alternative would be to consider motion on group manifolds such as $SU(2)$ or the Heisenberg–Weyl group, both analyzed in \cite{Balasubramanian:2022tpr}. However, these examples do not adequately capture the physics relevant for black holes. In the $SU(2)$ case, the correlator $C(t)$, and consequently the Krylov entropy $S_K$, remains bounded because $SU(2)$ admits only finite-dimensional representations. A similar qualitative limitation appears in the Heisenberg–Weyl example.\\

\noindent
Nevertheless, if we formally continue the $SL(2,\mathbb R)$ result while disregarding the overall factor of $i$, the resulting behaviour implies
$S_K \propto t$,
namely linear growth of Krylov entropy at intermediate times. This is precisely the behaviour expected for a black hole prior to the onset of late-time saturation effects.

\subsection{Approximate Krylov bases}

In \cite{Haque:2022ncl}, a scheme to approximate the Lanczos coefficients (by finding an appropriate 
approximation to the Krylov basis vectors) was devised.  Before proceeding, we provide a simple, 
but general corollary to those results.  These general considerations will be important when we study our toy model perturbatively.   \\ \\
Suppose that we are able to find a component of the true 
Krylov basis, i.e., a family of states $|K_{n}^{(0)}\rangle$ such that
\begin{equation}
|K_n\rangle = \cos\theta_n |K_n^{(0)} \rangle + \sin\theta_n |\tilde{K}_n\rangle,   \label{AppCond1}
\end{equation}
and for which 
\begin{equation}
\langle K_{n+1}^{(0)}| H | \tilde{K}_n\rangle = 0.    \label{AppCond2}
\end{equation}
The projection of the Hamiltonian onto the states $|K_n^{(0)}\rangle$ gives rise to a tri-diagonal matrix 
$$    \langle K_{n+1}^{(0)} | H | K_{n}^{(0)}\rangle = b^{(0)}_{n+1}.$$
The coefficients $b_{n+1}^{(0)}$ are not the Lanczos coefficients of the Hamiltonian, but rather approximate Lanczos coefficients.  We would like to relate these approximate coefficients to the true coefficients.   \\ \\ 
These expressions imply that 
\begin{eqnarray}
\langle K_{n+1}^{(0)}| H |K_n\rangle & = & \cos\theta_{n+1} b_{n+1}     \nonumber   \\
\langle K_{n+1}^{(0)}| H |K_n\rangle & = & \cos\theta_n \langle K_{n+1}^{(0)}| H |K_n^{(0)}\rangle\;, 
  \nonumber
\end{eqnarray}
so that the true Lanczos coefficients are related to the approximate Lanczos coefficients as 
\begin{equation}
b_{n+1} = \frac{\cos\theta_n}{\cos\theta_{n+1}} b_{n+1}^{(0)}.
\end{equation}
This general corollary relates the approximate Lanczos coefficients to the true Lanczos coefficients.  
Finding the $\cos\theta_n$ is still a difficult problem, but the above allows for sensible approximations.  
For example, in bosonic systems we would expect the $\cos(\theta_n)$ to be a decreasing function of $n$.  We 
thus expect the approximate Lanczos coefficients to be an underestimate of the true Lanczos 
coefficients in most reasonable cases.  See e.g. \cite{Haque:2022ncl} where this is shown for Hamiltonians from the Jacobi algebra.

\section{Perturbative and numerical studies}

In this section we study the early-time behaviour of the Krylov complexity and Krylov entropy in our toy model using both perturbative and numerical methods. Before turning to these calculations, however, it will be useful to briefly discuss the Izuka-Polchinski (IP) model and its relation to the present setup.

\subsection{IP and IOP model and our version for it}

In \cite{Iizuka:2008hg}, the IP model was introduced as a simplified framework intended to capture certain aspects of black-hole dynamics. The model consists of a Hermitian matrix degree of freedom $X_{ij}(t)$ transforming in the adjoint representation of $U(N)$, together with a complex vector $\phi_i(t)$ in the fundamental representation. Both fields possess harmonic-oscillator-type mass terms, and interact through an additional coupling between the adjoint and fundamental sectors.\\

\noindent
From the perspective adopted here, it is natural to interpret the matrix variable $X_{ij}(t)$ as describing the black-hole degrees of freedom, while the field $\phi_i(t)$ plays the role of radiation interacting with the background. In \cite{Iizuka:2023pov}, however, this interpretation is not made explicit, largely because the resulting behaviour of the Krylov entropy does not fully match the expectations for black-hole information dynamics. Instead, the authors interpret the matrix sector as describing a system of $N$ D0-branes, while $\phi_i$ is viewed as a probe D0-brane moving in the background generated by the matrix degrees of freedom. Using $X,\Pi$ for the black hole phase space variables and $\phi,\pi$ for the radiation phase space variables, the Hamiltonian reads
\be
H=\frac{1}{2}\Tr \Pi^2+\frac{m^2}{2}\Tr X^2 +\pi^\dagger \pi+M^2\phi^2\phi+\frac{g}{M}\left(\pi^\dagger X 
\pi+M^2\phi^\dagger 
X \phi\right)\;,
\ee
with $[X(t),\Pi(t')]=i\delta(t-t')$, $[\phi(t),\pi(t')]=i\delta(t-t')$ 
and $g=g_{YM}$, as expected for the YM theory on the D0-branes. The action is thus
\be
S=\int dt\Tr\left[\frac{\dot X^2}{2}-m^2 \frac{X^2}{2}\right]
+\int dt \left[\dot \phi^\dagger\dot \phi -M^2\phi^\dagger \phi
-\frac{g}{M}\left(\dot \phi^\dagger X\dot\phi+M^2\phi^\dagger X\phi\right)\right].
\ee
Our toy model differs from the IP model in several important respects. First, the radiation field $\phi$ transforms in the fundamental representation rather than the adjoint. As a consequence, the interaction term is quadratic in $\phi$ and linear in $X$, whereas in our model the interaction is cubic in the matrix variables and linear in $\phi$, reflecting the different index structure of the adjoint coupling. Second, the IP model contains only a single matrix degree of freedom $X$, whereas our construction retains three matrix scalars, as required by the mini-BMN structure. Finally, the IP model lacks intrinsic self-interactions among the matrix degrees of freedom. Taken together, these differences explain why the IP model does not fully capture the features we wish to study, particularly those associated with black-hole-like information dynamics and the behaviour of Krylov entropy.\\

\noindent
The IOP model \cite{Iizuka:2008eb} modifies the interaction structure of the original IP model by replacing the matrix field $MX_{ij}$ appearing between $\phi_i^\dagger$ and $\phi_j$ with the bilinear operator
$A^\dagger_{ik}A_{kj}$,
where $A$ and $A^\dagger$ are the annihilation and creation operators associated with the matrix degree of freedom $X_{ij}$. These are defined by
\be
A_{ij}
=
\frac{1}{\sqrt{2m}}
(\Pi_{ij}-imX_{ij}),
\qquad
A^\dagger_{ij}
=
\frac{1}{\sqrt{2m}}
(\Pi_{ij}+imX_{ij}).
\ee
Similarly, one may introduce oscillator variables for the fundamental field $\phi_i$,
\be
a_i
=
\frac{\pi_i^\dagger-iM\phi_i}{\sqrt{2M}},
\qquad
a_i^\dagger
=
\frac{\pi_i-iM\phi_i^\dagger}{\sqrt{2M}}.
\ee
In terms of these operators, the $\pi$- and $\phi$-dependent part of the Hamiltonian takes the form
\be
M(a^\dagger a+aa^\dagger)
+
g(a^\dagger Xa+aX^Ta^\dagger).
\ee
Thus the interaction is rewritten directly in terms of oscillator excitations of the matrix degrees of freedom and the probe sector.

\subsection{Lanczos algorithm from a perturbative two-point function from the IP model}
\label{sec:perturbativeip}

Since perturbative two-point functions are not commonly used in the literature as a direct route to computing Lanczos coefficients, we include in this section a brief case study illustrating the method as a proof of concept. In \cite{Iizuka:2023pov}, the authors analyzed the spectral density associated with a particular two-point function in the IP model. Their result was obtained numerically by solving the corresponding recursive relation
\begin{equation}
    \tilde G(T,\omega - m)  - \frac{4}{\nu_T^2}\frac{1}{ \tilde G(T,\omega)} + e^{- m/T} \tilde G(T,\omega + m) 
    = \frac{4 i \omega}{\nu_T^2}\,,
\end{equation}
obtained from the integration of a Dyson-Schwinger equation. Here, $\nu_T^2 = \nu^2/(1- e^{- m/T})$, 
$\nu$ is related to the coupling of the IP model, and $\tilde G(T,\omega)$ is a time-ordered thermal 
two-point function in frequency space. 
The moments $M_n$ can be found from the integrals
\begin{equation}
    M_{n}=\int_{-\infty}^{\infty}\frac{d\omega}{2\pi}\,\omega^{n}f(\omega),
\end{equation}
and, as explained in their paper, the spectral density $f(\omega)$ can be obtained from $\tilde G(T,\omega)$,
\begin{equation}
    f(\omega)=2\Re(\tilde G(T,\omega))\,.
\end{equation}
A finite set of 
moments and Lanczos coefficients can then be obtained by numerically integrating the spectral density. Here we show how the coefficients they obtained via the non-perturbative method can be reproduced from a perturbative result. We start by rewriting the recursive relation as
\begin{equation}
    \tilde G(T,\omega)=\left(-i \omega+\frac{\nu_T^2}{4}\left(\tilde G(T,\omega - m) + e^{- m/T} 
    \tilde G(T,\omega + m) \right)\right)^{-1}.
\end{equation}
Then, by recursively substituting $\tilde G(T,\omega)$ onto itself a number of times, we obtain something 
similar to a perturbative expansion. We perform a cut by finally substituting $\tilde G(T,\omega-m)$ and $\tilde 
G(T,\omega+m)$ by $\tilde G_0(T,\omega-m)$ and $\tilde G_0(T,\omega+m)$ on the right side (here we 
mean $\tilde G_0(T,\omega)\equiv(-i\omega)^{-1}$). For example, with zero substitutions, we have,
\begin{equation}
    \tilde G(T,\omega)=\left(-i \omega+\frac{\nu_T^2}{4}\left(\tilde G_0(T,\omega - m) 
    + e^{- m/T} \tilde G_0(T,\omega + m) \right)\right)^{-1},
\end{equation}
which we shall call a first order result (we're just introducing a convenient nomenclature 
fixed by the number of substitutions).\\

\noindent
Having obtained perturbative results for the spectral density $f(\omega)$, 
we integrate it using the Sokhotski-Plemelj-Fox formula,
\begin{equation}
    \label{eq_fox}
    \int_{a}^{b} dx \frac{h(x)}{(x+i\epsilon -m)^{n+1}} = \#\int_{a}^{b} dx \frac{h(x)}{(x-m)^{n+1}} 
    - i\pi \frac{h^{(n)}(m)}{n!},
\end{equation}
where $\#$ indicates the Hadamard finite part integral (analogous to the principal value integral 
in the case of a simple pole). However, since $f(\omega)=2\Re(\tilde G(T,\omega))$, we can 
simply integrate the real part of $\tilde G(T,\omega)$ or, conversely, we only need to keep 
the real part of the result of the integral. Since $\tilde G(T,\omega)$ is purely imaginary, we 
end up up keeping only the rightmost term in eq. \ref{eq_fox}, representing the residue of $h(x)$. 
This then gives us a certain number of moments and Lanczos coefficients which can be compared 
to the non perturbative ones.\\

\begin{figure}[hbt!]
    \centering
    \includegraphics[width=0.8\textwidth]{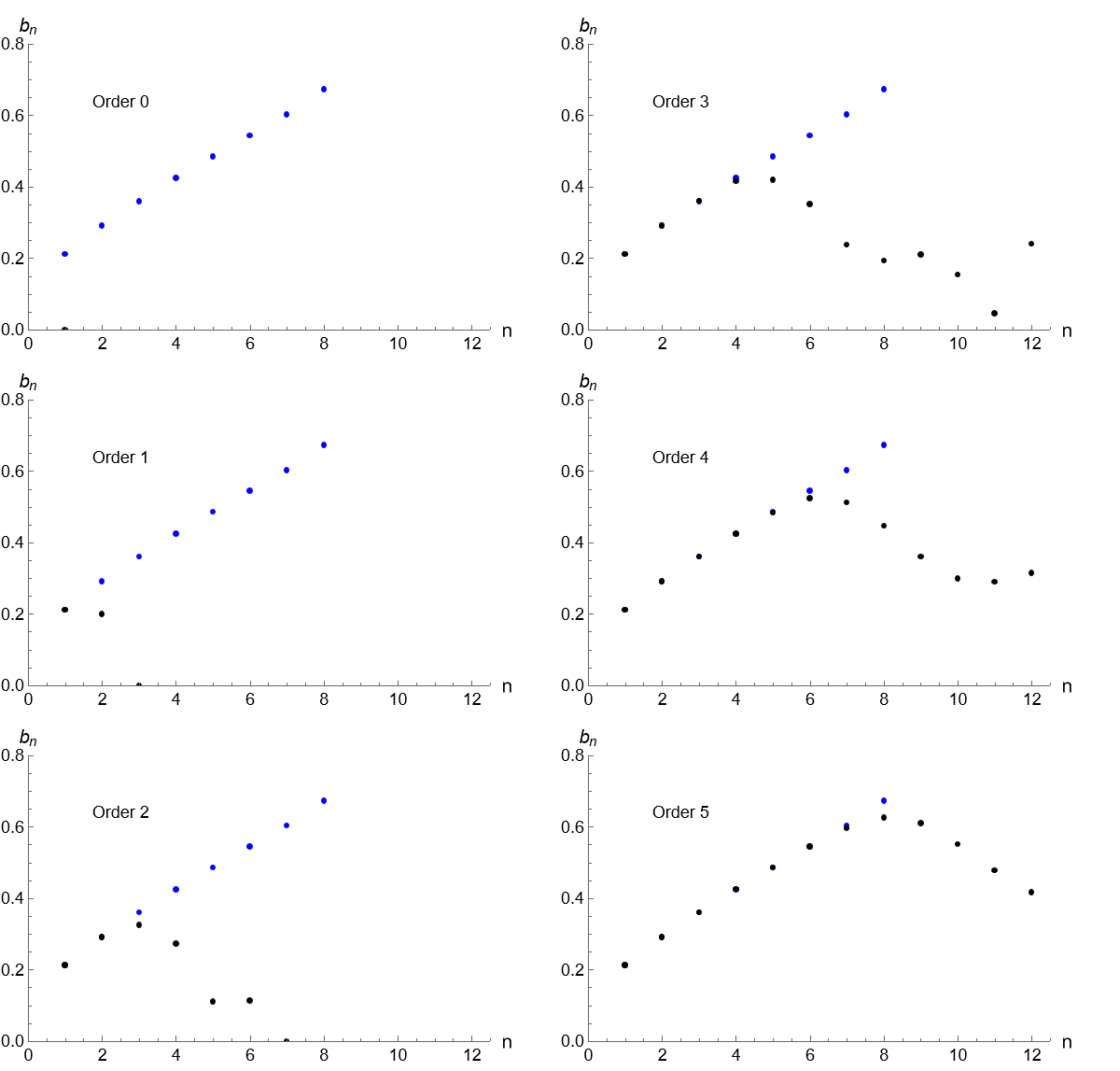}
    \caption{Comparison between the $b_n$ coefficients found with perturbative (black) and non-perturbative 
    methods (blue) for the 2-point function of the IP model. The perturbative results range from order 0 to 5 
    and the parameters are $m=2/10$, $\nu_T=3/10$, $T\to\infty$.}
    \label{fig:ordersbnip}
\end{figure}

\begin{figure}[hbt!]
    \centering
    \includegraphics[width=0.6\textwidth]{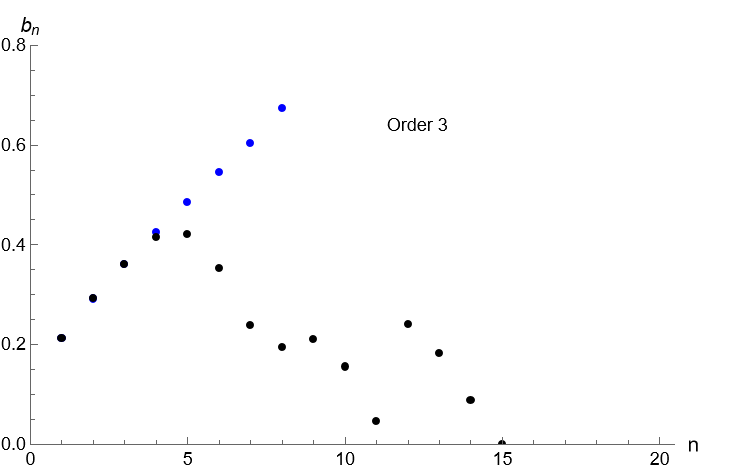}
    \includegraphics[width=0.6\textwidth]{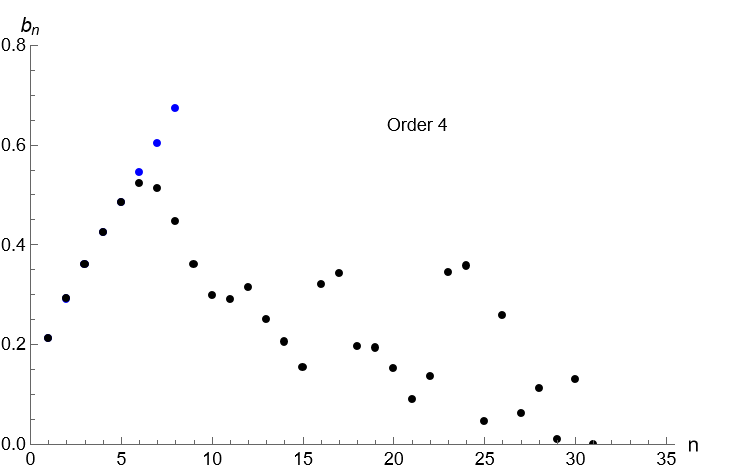}
    \includegraphics[width=0.6\textwidth]{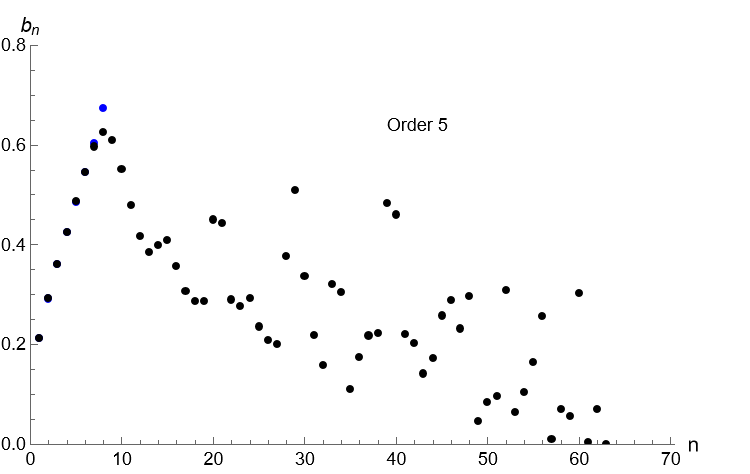}
    \caption{Comparison between the $b_n$ coefficients found with perturbative (black) and non-perturbative 
    methods (blue) for the 2-point function of the IP model. The perturbative results range from order 3 to 5 
    and the parameters are $m=2/10$, $\nu_T=3/10$, $T\to\infty$.}
    \label{fig:order345bnip}
\end{figure}

\begin{figure}[hbt!]
    \centering
    \includegraphics[width=0.8\textwidth]{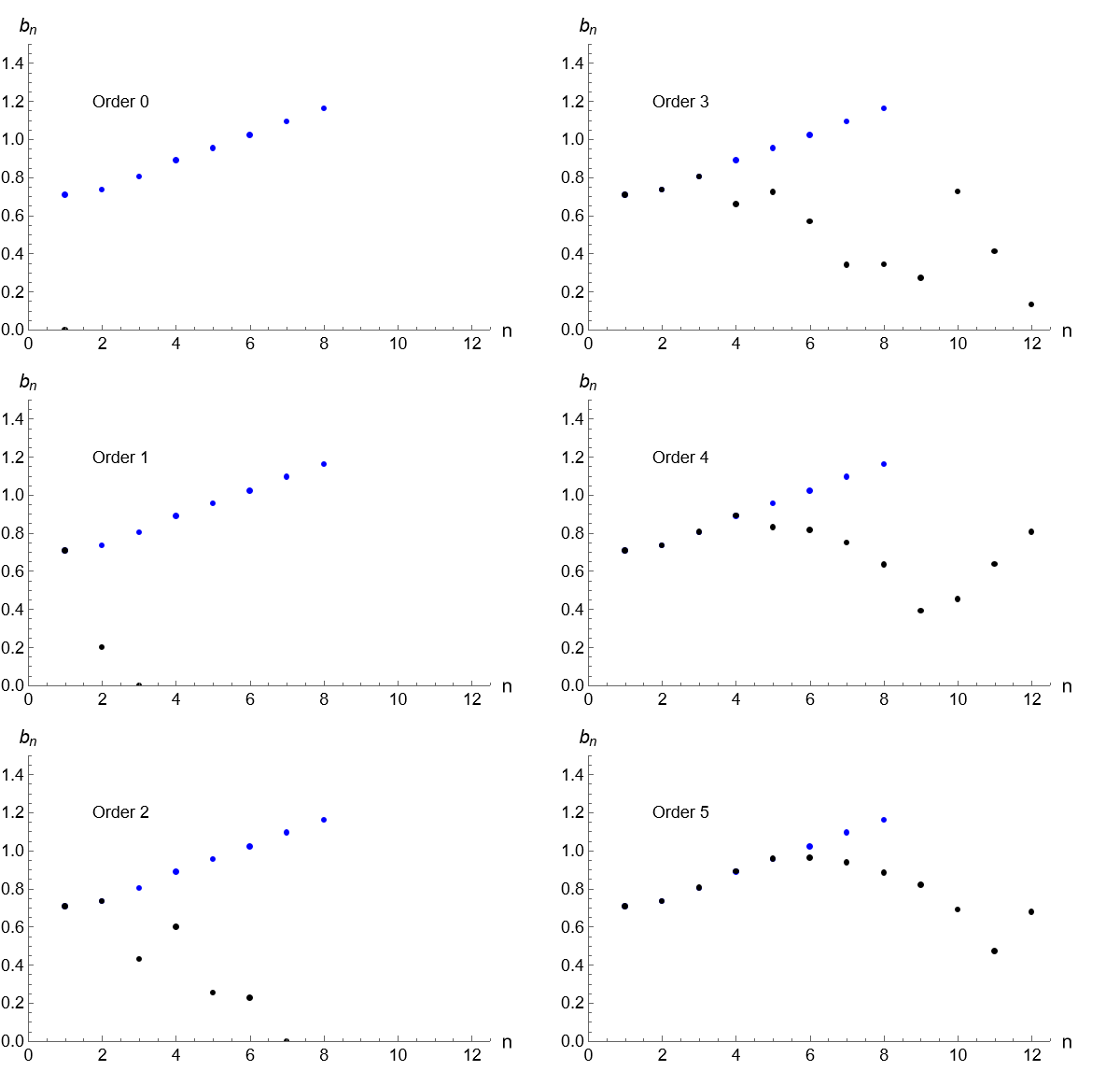}
    \caption{Comparison between the $b_n$ coefficients found with perturbative (black) and non-perturbative 
    methods (blue) for the 2-point function of the IP model. The perturbative results range from order 0 to 5 
    and the parameters are $m=2/10$, $\nu_T=1$, $T\to\infty$.}
    \label{fig:ordersbnipfull}
\end{figure}

\begin{figure}[hbt!]
    \centering
    \includegraphics[width=0.6\textwidth]{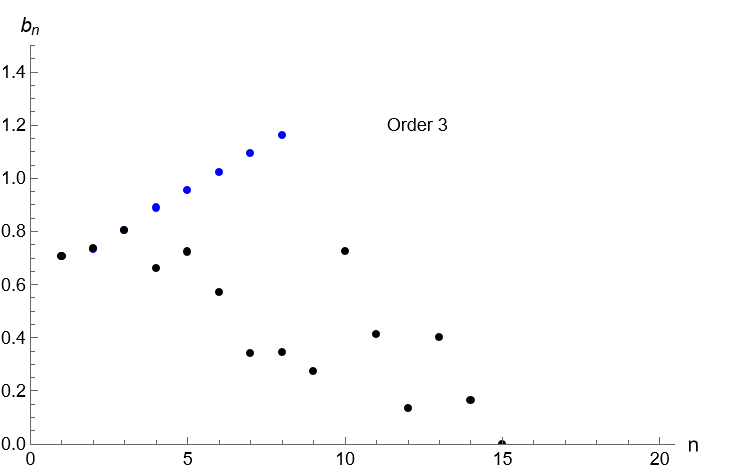}
    \includegraphics[width=0.6\textwidth]{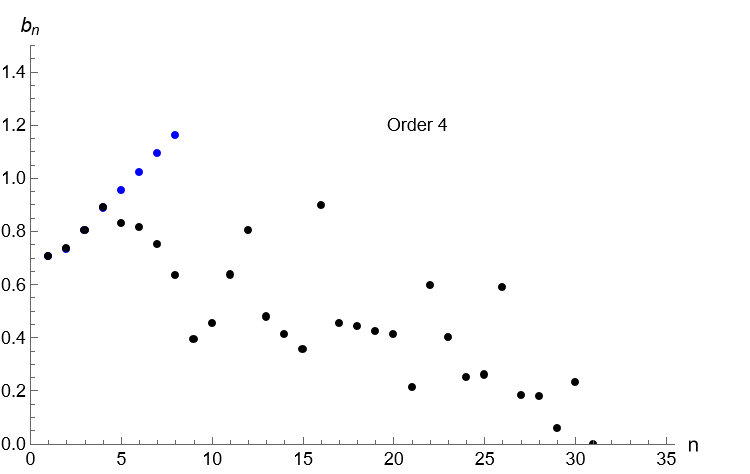}
    \includegraphics[width=0.6\textwidth]{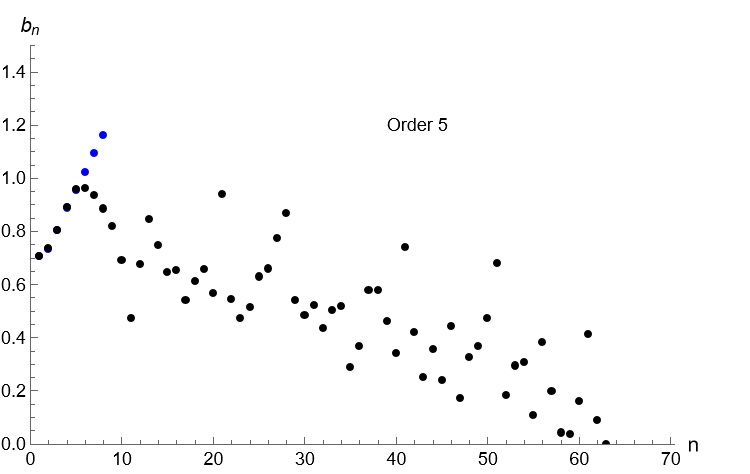}
    \caption{Comparison between the $b_n$ coefficients found with perturbative (black) and non-perturbative 
    methods (blue) for the 2-point function of the IP model. The perturbative results range from order 3 to 5 
    and the parameters are $m=2/10$, $\nu_T=1$, $T\to\infty$.}
    \label{fig:order345bnipfull}
\end{figure}

\noindent
We've obtained the Lanczos coefficients in both ways for the case $m=2/10$, $\nu_T=3/10$, $T\to\infty$. 
The perturbative results for the coefficients $b_n$ (black dots) of different orders are compared to the non-perturbative ones (blue dots) in Figures \ref{fig:ordersbnip}, \ref{fig:order345bnip}. We omit the $a_n$ 
coefficients, because they were always found to be $0$ and matching in number the amount of Krylov 
dimensions established by the $b_n$'s.
The perturbative results are seen to agree remarkably well with the corresponding nonperturbative calculations. As expected, obtaining accurate Lanczos coefficients at progressively larger values of $n$ requires carrying the perturbative expansion to increasingly high orders. An interesting feature is that the effective dimension of the Krylov basis obtained perturbatively is itself controlled by the order to which the perturbation series is truncated. A similar behavior is found by analysis of the $
\nu_T=1$ case, with results given in Figures \ref{fig:ordersbnipfull}, \ref{fig:order345bnipfull}.

\FloatBarrier
\subsection{Dyson-Schwinger equations method}
\label{sec:perturbative toy model}

The Dyson–Schwinger equation provides a diagrammatic reformulation of perturbation theory in the large-$N$ limit, where only planar diagrams contribute. In favorable cases, most notably in models such as SYK, these equations may be solved exactly. More generally, however, they should be viewed as an efficient organizational framework for resumming perturbative contributions and extracting approximate solutions.\\

\noindent
For the toy-model Hamiltonian \eqref{toyH} describing the black-hole sector, the regime
$\omega \ll T$
allows us to treat the matrix fields $X^I$ as effectively free from self-interactions, as discussed in section \ref{sec:BMNtoymodel}. The intrinsic BMN interaction terms may therefore be neglected to leading order, although they could in principle be reintroduced systematically as perturbative corrections. In this approximation, the only remaining interaction is the cubic coupling
$\phi XXX$. Moreover, in order for the radiation field $\phi$ to define a well-behaved oscillator sector, it is necessary to include a mass term for $\phi$. We discuss this issue in greater detail in section \ref{sec:towards states}, but for the perturbative analysis presented here we incorporate this modification from the outset.
Thus, for the purposes of this section, the effective toy-model action takes the form
\begin{equation}
\begin{aligned}
S=\frac{N}{\lambda}\int dt\; \Tr \biggl\{\sum_{I=1}^3\left[\frac{1}{2}(\dot X^I)^2
-\frac{(2\omega)^2}{2}(X^I)^2\right]
&+\left[\frac{1}{2}(\dot \phi)^2 -\frac{(\omega_\phi)^2}{2}(\phi)^2\right]\\
&-i(2\omega +\tilde g \phi)\sum_{I,J,K=1}^3\epsilon_{IJK}X^I X^J X^K \biggl\}\;.
\end{aligned}
\end{equation}
We've found the Dyson-Schwinger equation for a 2-point function of X's,
\begin{align}
\begin{split}
    \label{eq_ds final}
    \expect{\mathcal{T}  &X_{\eta\gamma}^A (\ctau_2) X_{\nu\mu}^T (\ctau_1)}_{\beta} =
    \Delta_{0X}(\tau_2,\tau_1) \delta_{\mu\eta}\delta_{\gamma\nu}\delta^{AT} \\
    &- i\Tilde{g}\epsilon_{IJK}\delta^{IT} \sum_{i,j} \int_{0}^{\beta} d\ctau_h \Delta_{0X}(\tau_1,\tau_h)
    \Bigg[ \expect{ \mathcal{T}  X_{\eta\gamma}^A(\ctau_2) X_{\nu i}^J (\ctau_h) X_{i j}^K (\ctau_h) 
    \phi_{j \mu} (\ctau_h)}_{\beta} \\
    &+ \expect{ \mathcal{T}  X_{\eta\gamma}^A(\ctau_2) X_{\nu i}^J (\ctau_h) \phi_{i j} (\ctau_h) 
    X_{j \mu}^K (\ctau_h)}_{\beta}
    + \expect{ \mathcal{T}  X_{\eta\gamma}^A(\ctau_2) \phi_{\nu i} (\ctau_h) X_{i j}^J (\ctau_h) 
    X_{j \mu}^K (\ctau_h)}_{\beta} \Bigg],
\end{split}
\end{align}
where $\ctau\equiv-i\tau$, $\Delta_{0X}$ and $\Delta_{0\phi}$ are free propagators for the oscillators $X$ 
and $\phi$ ($\Delta_{0X} (\tau_a,\tau_b) \equiv \expect{\mathcal{T} X_{ij}^I(\ctau_a)X_{ji}^I(\ctau_b)}
_{\beta,0}$), and we've made explicit the indices of the adjoint representation of $SU(N)$.
The terms on the right side of eq. \ref{eq_ds final} can be represented diagrammatically:
\begin{align}
    \begin{split}
        &\Delta_{0X}(\tau_2,\tau_1)\delta_{\mu\eta}\delta_{\gamma\nu}\delta^{AT} \\
        &= \quad \includegraphics[valign=c,scale=.4]{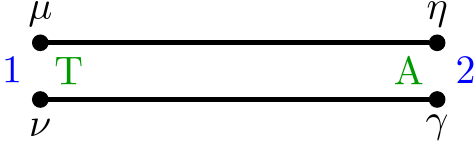} \;,
    \end{split}
    \\ \nonumber
    \\
    \begin{split}
        &-i (\tilde{g}) \epsilon_{IJK}\delta^{IT} \sum_{i,j} \int_{0}^{\beta} d\tau_h \Delta_{0X}(\tau_1,\tau_h)
        \expect{\mathcal{T}  X_{\eta\gamma}^A(\ctau_2) X_{\nu i}^J (\ctau_h) X_{i j}^K (\ctau_h) 
        \phi_{j \mu} (\ctau_h)}_{\beta} \\
        &= \quad \includegraphics[valign=c,scale=.4]{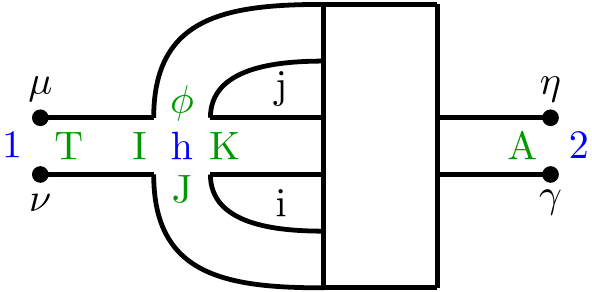} \;.
    \end{split}
\end{align}
They can be further simplified in the following way,
\begin{equation}
    \includegraphics[valign=c,scale=.4]{pvertex} \equiv 
    \includegraphics[valign=c,scale=.4]{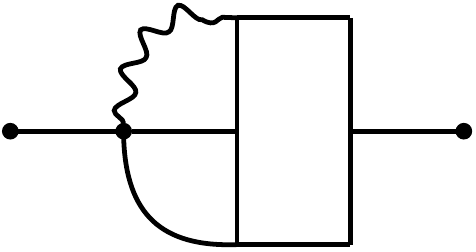} \;,
\end{equation}
so that eq. \ref{eq_ds final} may be written as
\begin{align}
\begin{split}
    \label{eq_x2 first}
    &\includegraphics[valign=c,scale=.4]{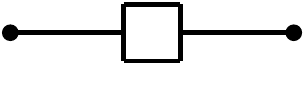} = \includegraphics[valign=c,scale=.4]{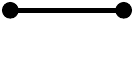} \\
    &+ \includegraphics[valign=c,scale=.4]{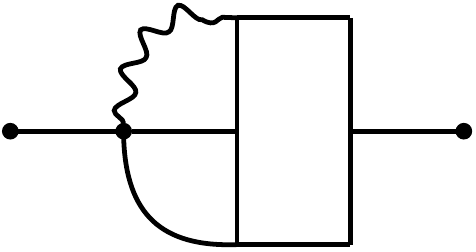} + \includegraphics[valign=c,scale=.4]{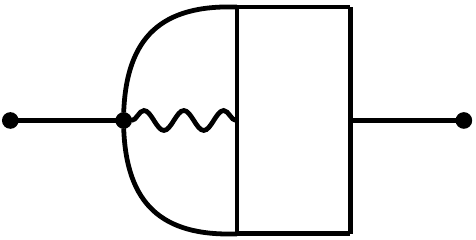} + \includegraphics[valign=c,scale=.4]{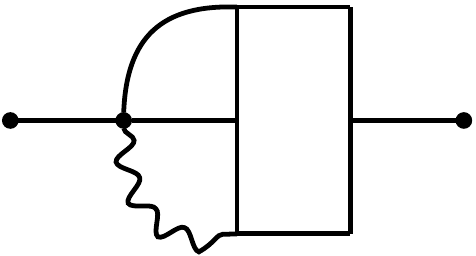} \;.
\end{split}
\end{align}
One finds that the two-point functions are always given in terms of higher n-point functions, 
represented here by boxes. For example, with two vertices,
\begin{align}
\begin{split}
    \label{eq_x2 first-2}
    &\includegraphics[valign=c,scale=.4]{x2} = \includegraphics[valign=c,scale=.4]{x2free} \\
    &+ \includegraphics[valign=c,scale=.4]{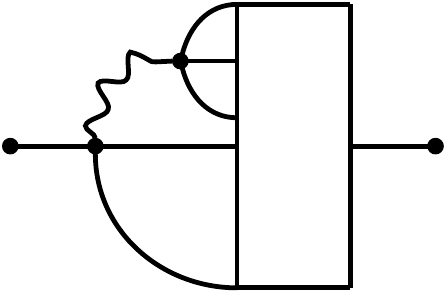} + \includegraphics[valign=c,scale=.4]{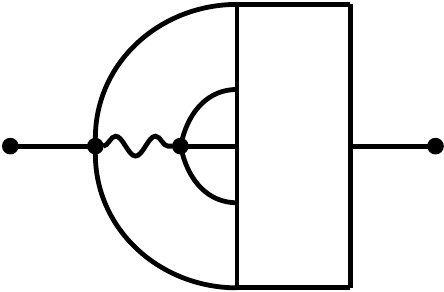} + \includegraphics[valign=c,scale=.4]{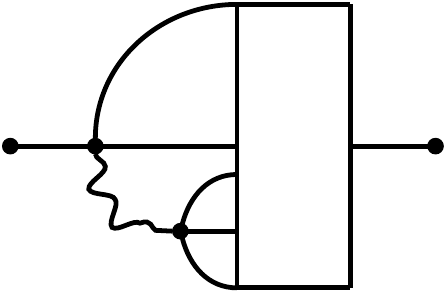} \;,
\end{split}
\end{align}
so it's not possible to find a closed equation from which one could extract numerical or analytical non-perturbative solutions. We're left with the possibility of using perturbation theory: we cut the 
series at some point (by considering the top $n$-point functions as free) and solve for the lower $n$-point functions.
Up to second order, the perturbative two-point function is written as (keeping only planar diagrams),
\begin{align}
\begin{split}
    &\includegraphics[valign=c,scale=.4]{x2} = \includegraphics[valign=c,scale=.4]{x2free} \\
    &+ \includegraphics[valign=c,scale=.4]{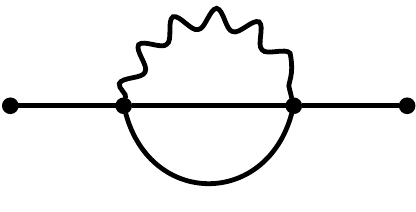} + \includegraphics[valign=c,scale=.4]{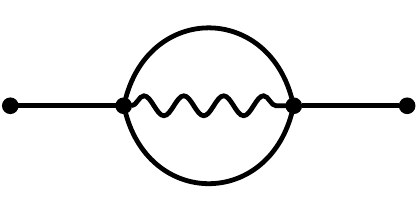} + \includegraphics[valign=c,scale=.4]{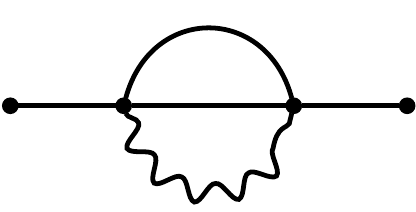}\;,
\end{split}
\end{align}
so the diagram that matters to us is
\begin{align}
    &\includegraphics[valign=c,scale=.4]{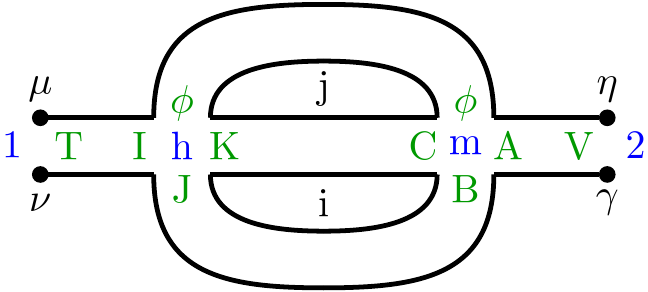} \nonumber \\
\begin{split}
     &= (-i)^2 ( N\tilde{g})^2 \delta_{\mu\eta}\delta_{\nu\gamma}\epsilon_{IJK}\epsilon_{ABC}
     \delta^{TI}\delta^{AV} \delta^{JB} \delta^{KC} \\
    &\cdot \int_0^{\beta}\int_0^{\beta} d\tau_h d\tau_m \Delta_{0X} (\tau_1,\tau_h) \Delta_{0X} (\tau_m,\tau_2)
    \Delta_{0\phi} (\tau_h,\tau_m) (\Delta_{0X} (\tau_h,\tau_m) )^2
\end{split}
\\
\begin{split}
     &= (-i)^2 ( N\tilde{g})^2 \delta_{\mu\eta}\delta_{\nu\gamma}\epsilon_{IJK}\epsilon_{ABC}
     \delta^{TI}\delta^{AV} \delta^{JB} \delta^{KC} \\
    &\cdot \int_0^{\beta}\int_0^{\beta} d\tau_h d\tau_m \Delta_{0X} (\tau_1-\tau_h) \Delta_{0X} (\tau_m-\tau_2)
    \Delta_{0\phi} (\tau_h-\tau_m) (\Delta_{0X} (\tau_h-\tau_m) )^2,
\end{split}
\end{align}
and the two-point function itself
\begin{align}
\begin{split}
    &\expect{\mathcal{T}  X_{\eta\gamma}^A (\ctau_2) X_{\nu\mu}^T (\ctau_1)}_{\beta} \\
    & =\Delta_{0X}(\tau_2-\tau_1) \delta_{\mu\eta}\delta_{\gamma\nu}\delta^{AT} + 3(-i)^2 ( N\tilde{g})^2 \delta_{\mu\eta}\delta_{\nu\gamma}\epsilon_{IJK}\epsilon_{ABC}
     \delta^{TI}\delta^{AV} \delta^{JB} \delta^{KC} \\
    &\cdot \int_0^{\beta}\int_0^{\beta} d\tau_h d\tau_m \Delta_{0X} (\tau_1-\tau_h) \Delta_{0X} (\tau_m-\tau_2)
    \Delta_{0\phi} (\tau_h-\tau_m) (\Delta_{0X} (\tau_h-\tau_m) )^2.
\end{split}
\end{align}
\noindent
Let us rewrite this expression as
\begin{align}
\begin{split}
    \label{eq_correlator tau}
    &\Delta_X(\tau_2-\tau_1) = \Delta_{0X} (\tau_2-\tau_1) \\
    &+ \Lambda \int_0^{\beta}\int_0^{\beta} d\tau_h d\tau_m \Delta_{0X} (\tau_1-\tau_h) \Delta_{0X} (\tau_m-\tau_2)
    \Delta_{0\phi} (\tau_h-\tau_m) (\Delta_{0X} (\tau_h-\tau_m) )^2,
\end{split}
\end{align}
\noindent
where $\Lambda$ is a constant and $\Delta_X$ denotes the interacting propagator up to second order. We perform Fourier transforms and introduce Matsubara frequencies, denoting them by $\Omega$ instead of the usual lowercase $\omega$ to avoid confusion with the parameters of our model,
\begin{align}
    \Delta(i\Omega) &= \int_0^\beta d\tau e^{(i\Omega \tau)} \Delta(\tau) \\
    \Delta (\tau) &= \frac{1}{\beta} \sum_n e^{(-i\Omega \tau)} \Delta(i\Omega), \quad \Omega = \frac{2\pi n}{\beta}.
\end{align}
For a free oscillator of frequency $2\omega$ the propagator is
\begin{equation}
    \Delta_0(i\Omega) = \frac{1}{\Omega^2 + (2\omega)^2},
\end{equation}
and note that we emphasize $2\omega$ instead of the usual $\omega$ because the frequency of our oscillators $X$ is indeed defined as $2\omega$.
Our two-point function becomes
\begin{align}
\begin{split}
    &\Delta_X(i\Omega_1) = \Delta_{0X} (i\Omega_1)\\
    &+ \frac{\Lambda}{\beta^2} \sum_{\Omega_2,\Omega_3}
    (\Delta_{0X} (i\Omega_1))^2 \Delta_{0X} (i\Omega_2) \Delta_{0X} (i\Omega_3)
    \Delta_{0\phi} (i(\Omega_1-\Omega_2-\Omega_3)),
\end{split}
\end{align}
\noindent
or, writing the propagators explicitly,
\begin{align}
\begin{split}
\label{eq_to solve}
    &\Delta_X(i\Omega_1)= \frac{1}{\Omega_1^2 + (2\omega)^2} \\
    &+ \frac{\Lambda}{\beta^2} \sum_{\Omega_2,\Omega_3}
    \left(\frac{1}{\Omega_1^2 + (2\omega)^2} \right)^2 \left(\frac{1}{\Omega_2^2 + (2\omega)^2} \right) \left(\frac{1}{\Omega_3^2 + (2\omega)^2} \right)
    \left(\frac{1}{(\Omega_1-\Omega_2-\Omega_3)^2 + (\omega_\phi)^2} \right).
\end{split}
\end{align}
Upon solving these frequency sums analytically with the usual residue method, we obtain an expression for $\Delta_X(i\Omega)$. This is related to a retarded propagator in frequency space ($\tilde{C}_R(\Omega)$) via analytical continuation,
\begin{equation}
    \tilde{C}_R(\Omega) = -i\Delta_X(\Omega + i\epsilon),
\end{equation}
and from this retarded propagator we obtain the one we actually need ($\tilde{C}^+(\Omega)$), in order to calculate the moments ($M_n$) and the Lanczos coefficients,
\begin{align}
    C^+(t)&\equiv\expect{ X_{\eta\gamma}^A (t) X_{\nu\mu}^T (0)}_{\beta},\\
    \tilde{C}^+(\Omega)&\equiv \int_{-\infty}^{\infty}dt \; e^{i\Omega t} C^+(t),\\
    \tilde{C}^+(\Omega) &= \frac{2\Re (\tilde{C}_R(\Omega))}{1-e^{-\beta\Omega}},\\
    M_n&=\int_{-\infty}^{\infty} \frac{d\Omega}{2\pi}\Omega^n \tilde{C}^+(\Omega).
\end{align}
The moments are found via integration using the Sokhotski-Plemelj-Fox formula (eq. \ref{eq_fox}), as described in Section \ref{sec:perturbativeip}.
When using the following set of parameters,
\begin{equation}
    N = 10^{20},\quad
    (2\omega) = 1,\quad
    \omega_{\phi}=4,\quad
    T = 10^3,\quad
    \tilde{g}= 10^{-30},\quad
    \lambda = N,
\end{equation}
which satisfy
\begin{align}
\begin{split}
    \frac{T}{(2\omega)}&=10^{3}\gg1,\\
    \frac{\lambda^{1/3}}{T}&=4641.59\gg1,\\
    N^{1/3}\frac{T}{\lambda^{1/3}}&=10^3\gg1,
\end{split}
\end{align}
we obtain the Lanczos coefficients given in Table \ref{tab:lanczosdyson}. With only eleven dimensions 
in the Krylov basis and $b_n$'s varying wildly (besides some being imaginary), this result fails to give 
us any clue about the real growth pattern of coefficients for this model. From the results given in 
Section \ref{sec:perturbativeip}, we infer that we would need many more orders of perturbation 
to obtain something useful.  \\ \\
In the subsequent sections, we will detail an alternative approach (that is perturbative in nature) for tackling the early-time behavior of our model.  
\begin{table}[hbt!]
    \centering
    \begin{tabular}{c|c|c}
        $n$  &  $a_n$  &  $b_n$\\
        \hline
        $0$  &   $5.0000\cdot10^{-4}$  &  ---\\
        $1$  &   $-5.0000\cdot10^{-4}$  &  $1.0000$\\
        $2$  &   $6.0000\cdot10^3$  &  $5.3033\cdot10^{-11}$\\
        $3$  &   $-3.0000\cdot10^3$  &  $(4.8990\cdot10^3)i$\\
        $4$  &   $-3.0000\cdot10^3$  &  $(1.7321\cdot10^3)i$\\
        $5$  &   $1.1451\cdot10^{-2}$  &  $4.4887\cdot10^{-3}$\\
        $6$  &   $-1.0363\cdot10^{-2}$  &  $4.7856$\\
        $7$  &   $1.7067\cdot10^{-3}$  &  $1.4750$\\
        $8$  &   $-2.7945\cdot10^{-3}$  &  $2.3641$\\
        $9$  &   $5.0000\cdot10^{-4}$  &  $(9.8456\cdot10^{-8})i$\\
        $10$  &   $-5.0000\cdot10^{-4}$  &  $1.0000$
    \end{tabular}
    \caption{Lanczos coefficients obtained for the BMN toy model from a 
    second order approximation to a two-point function.}
    \label{tab:lanczosdyson}
\end{table}

\FloatBarrier
%\subsection{Calculations from the spectral distribution}

\subsection{Constructing states for our toy model}

In order to understand the evolution of the (Krylov) complexity, it is useful to understand the Hilbert space
of states of our toy model. 

\subsubsection{First try: simplified version}

Our toy model Hamiltonian is actually (\ref{toyH}), but for simplicity, and to understand how to 
proceed, we will first consider the model with a single parameter, the cubic coupling $g$, and no $\phi$ 
field, so 
\be
H=\Tr\left\{\sum_{I=1}^ 3\left[\frac{(p^ I)^ 2}{2}+\frac{(X^ I)^ 2}{2}\right]+\sum_{I,J,K=1}^ 3 g \epsilon_{IJK} X^ I X^ J X^ K\right\}.
\ee
Quantize the system via the usual
\be
[X^ I _{ij},p^ I_{kl}]=i\delta^ {IJ}\delta_{ik}\delta_{jl}.
\ee
Then define the creation/annihilation operator basis ($SU(N)$ matrix indices suppressed), 
\bea
a^ I&=&\frac{X^ I+ip^ I}{\sqrt{2}}\;,\;\;
a^ {\dagger I}=\frac{X^ I-ip^ I}{\sqrt{2}}\Rightarrow\cr
X^I&=&\frac{a^ I+a^ {\dagger I}}{\sqrt{2}}\;,\;\;
p^ I=\frac{a^ I-a^ {\dagger I}}{i\sqrt{2}}.
\eea
The (quantum) Hamiltonian becomes 
\bea
H&=& \sum_{I=1}^ 3\sum_{i,j=1}^ N\frac{a^I_{ij}a^ {\dagger I }_{ji}+a^ {\dagger I}_{ij}a^ I_{ji}}{2}
+\frac{g}{2\sqrt{2}}\sum_{I,J,K=1}^ 3\sum_{i,j,k=1}^ N\epsilon_{IJK}(a^ I_{ij}+a^ {\dagger I}_{ij})(a^ J_{jk}+a^ {\dagger J}_{jk})
(a^ K_{ki}+a^ {\dagger K}_{ki})\cr
    &=&H_0+gH_{\rm int}   \label{Hsimplified}
\eea
Thus the Hamiltonian splits into $H_0$, the free Hamiltonian (sum of free harmonic oscillators) and $gH_{\rm int}$, the 
cubic interaction piece. However the full Hamiltonian is invariant under $SO(3)\simeq SU(2)$ rotating the $I$'s. The generators of this symmetry are 
\be
J^ I=\sum_{J,K=1}^ 3\epsilon_{IJK}\Tr[X^ Jp^ K]=-\frac{i}{2}\sum_{J,K=1}^ 3\epsilon_{IJK}\sum_{i,j=1}^ N(a^ J_{ij}+a^ {\dagger J}_{ij})
(a^ K_{ji}-a^ {\dagger K}_{ji}).
\ee
One can check that, indeed, 
\be
[H,J^ I]=0\;,
\ee
so that the Hamiltonian is invariant, and also that 
\be
[J^ I,J^ J]=i\sum_{K=1}^3\epsilon^{IJK}J^ K\;,
\ee
so the symmetry generators obey the $SU(2)$ algebra. We can form the usual $J_\pm=J_1\pm i J_2, J_3$ basis, in terms of which 
\be
[J_3,J_\pm]=\pm J_\pm\;,\;\; [J_+,J_-]=2J_3.
\ee
We obtain
\bea
J_+&=& -i\Tr[a^ {\dagger 2}a^ 3-a^ {\dagger 3}a^ 2+ia^ {\dagger 3}a^ 1-ia^ {\dagger 1}a^ 2]\cr
J_-&=& -i\Tr[a^ {\dagger 2}a^ 3-a^ {\dagger 3}a^ 2-ia^ {\dagger 3}a^ 1+ia^ {\dagger 1}a^ 2]\cr
J_3&=&-i\Tr[a^ {\dagger 1}a^ 2-a^ {\dagger 2}a^ 1].
\eea
The $a^ I_{ij},a^ {\dagger I}_{ij}$ are used to construct an eigenbasis for the free hamiltonian $H_0$, which is not 
an eigenbasis of the full Hamiltonian, but it is useful as a starting point. The (unique) vacuum state
is $|0\rangle$, defined by 
\be
a^ I_{ij}|0\rangle=0\;,\forall I, i,j.
\ee
However, in constructing the higher energy states, one has to remember that we also have $SO(3)\simeq SU(2)$ invariance, 
which orders the states in terms of their $SU(2)$ representations, like $\{|j,m\rangle\}_m$ at fixed $j$. In the spin $j$ representation, the highest weight state is $|j,-j\rangle$, defined by 
\be
J_-|j,-j\rangle=0\;,\;\;
J_+|j,-j\rangle=\sqrt{2j}|j,-j+1\rangle\;,\;\;
J_3|j,-j\rangle=-j|j,-j\rangle.
\ee
It is easy then to see that the vacuum state $|0\rangle$ corresponds to the spin $j=0$ state, since that 
gives $J_+|0\rangle=J_-|0\rangle=J_3|0\rangle=0$. To construct the states in the representation $j=1$, we construct
\be
|j=1,-1\rangle_{kl}=\frac{1}{\sqrt{2}}(a^{\dagger 2}_{kl}+ia^ {\dagger 1}_{kl})|0\rangle.
\ee
Indeed, we see that 
\be
J_3|1,-1\rangle_{kl}=-|1,-1\rangle_{kl}\;,
J_+|1,-1\rangle_{kl}=\sqrt{2}ia^ {\dagger 3}_{kl}|0\rangle\equiv \sqrt{2}|1,0\rangle_{kl}\;,\;\;
J_-|1,-1\rangle_{kl}=0.
\ee
Then 
\be
|1,0\rangle_{kl}=ia^{\dagger 3}_{kl}|0\rangle\;,
\ee
since 
\be
J_3|1,0\rangle_{kl}=0\;,
\ee
and then finally 
\be
J_+|1,0\rangle_{kl}=(a^ {\dagger 2}_{kl}-ia^ {\dagger 1}_{kl})|0\rangle\equiv |1,+1\rangle_{kl}.
\ee
We, however, are only interested in $SU(N)$ gauge invariant states. For a single $a^ \dagger_{kl}$, that would have been the trace, 
except it vanishes for $SU(N)$. So the first nontrivial one is the state 
\be
|2,-2\rangle_{ii}=\frac{1}{2}\Tr[(a^ {\dagger 2}+ia^ {\dagger 1})^ 2]|0\rangle=
\frac{1}{2}(a^{\dagger 2}_{kl}+ia^ {\dagger 1}_{kl})(a^ {\dagger 2}_{lk}
+ia^{\dagger 1}_{lk})|0\rangle.
\ee
The descendent states are
\bea
|2,-1\rangle_{ii}&=&\frac{i}{\sqrt{2}}\Tr[(a^{\dagger 2}+ia^{\dagger 1})a^{\dagger 3}]|0\rangle=\frac{i}{\sqrt{2}}
(a^{\dagger 2}_{kl}+ia^{\dagger 1}_{kl})a^{\dagger 3}_{lk}|0\rangle\cr
|2,0\rangle_{ii}&=&-\Tr[(a^{\dagger 3})^2]|0\rangle=-a^{\dagger 3}_{kl}a^{\dagger 3}_{lk}|0\rangle\cr
|2,+1\rangle_{ii}&=&\frac{i}{\sqrt{2}}\Tr[(a^{\dagger 2}-ia^{\dagger 1})a^{\dagger 3}]|0\rangle 
=\frac{i}{\sqrt{2}}(a^{\dagger 2}_{kl}-ia^{\dagger 1}_{kl})a^{\dagger 3}_{lk}|0\rangle\cr
|2,+2\rangle_{ii}&=&\frac{1}{2}\Tr[(a^{\dagger 2}-ia^{\dagger 1})^2]|0\rangle
=\frac{1}{2}(a^{\dagger 2}_{kl}-ia^{\dagger 2}_{lk})(a^{\dagger 2}_{lk}-ia^{\dagger 1}_{lk})|0\rangle.
\eea
Note that because of the cyclicity of the trace (and the fact that the $a^\dagger$'s commute), the states belong to the symmetric
representation. \\

%{\em The only thing I am not sure about is whether we need to subtract the "trace" of the representation 
%(such that we get the symmetric traceless representation, naturally of spin 2), 
%\be
%\frac{1}{3}\Tr\left[\frac{1}{2}(a^{\dagger 2}+ia^{\dagger 1})^2-(a^{\dagger 3})^2+\frac{1}{2}(a^{\dagger 2}-ia^{\dagger 1})^2\right]|
%0\rangle\;,
%\ee
%out of all the states above. We would need to calculate the effect of the $J_+,J_-,J_3$ on them and check against the expectation.
%Note that the combination above has $J_3(...)=0$, so it doesn't change the $m$ of the state. We need then to check only the 
%$J_+$ and $J_-$ action on the state.

\noindent
Actually,  one needs to subtract the trace from the $m=0$ components, so for the above 
replace $|2,0\rangle_{ii}$ with (the formula below comes from the known relation between $r^lY_{lm}$ and the 
$x_1,x_2,x_3$ Cartesian coordinates on the sphere; note that the added term has still $J_3=0$)
\bea
|2,0\rangle_{ii}&=&-\Tr\left[(a^{\dagger 3})^2-\frac{1}{3}((a^{\dagger1})^2
+(a^{\dagger 2})^2+(a^{\dagger 3})^2)\right]|0\rangle\cr
&=&-\frac{1}{3}\left[2a^{\dagger3}_{kl}a^{\dagger 3}_{lk}-(a^{\dagger2}+ia^{\dagger 1})_{kl}(a^{\dagger2}-i
a^{\dagger 1})_{lk}\right]|0\rangle.
\eea
In order to calculate the Lanczos coefficients and the complexity, 
we could either take $|\psi\rangle $ to be the vacuum $|0\rangle$, 
or the above $|2,-2\rangle_{ii}$ state, and the 
${\cal O}_n$ to be $a^ {\dagger I}_{kl}$. One question remains: do we take only gauge invariant operators, or not? If not, we will calculate
\be
(a^ {\dagger I}_{ij}|a^ {\dagger J}_{kl})_\b =\frac{1}{Z}\Tr\left[e^ {-\b \hat H}a^I_{ij}a^ {\dagger J}_{kl}\right].
\ee

\subsubsection{Towards states for our toy model}
\label{sec:towards states}
 
We want to describe the states of our toy model Hamiltonian in (\ref{toyH}) in the end, 
but for that we need to put it in the $a,a^\dagger$ basis. That means that, with respect to the previous 
subsubsection, we need to:
\begin{itemize}
    \item introduce the $\omega$
    \item introduce the field $\phi$, coupling to the $X$'s. 
    \item we will see that we also need to put an (IR) regulator mass for $\phi$. 
\end{itemize}
\noindent
Indeed, we have 
\be
\phi_{ij}=\frac{a_{ij}^\phi+a^{\dagger\phi}_{ij}}{\sqrt{2\omega_\phi}}\;,
\ee
so introducing $\phi_{ij}$ in the Hamiltonian makes it singular, unless we have $\omega_\phi\neq 0$. However, in the free part of the Hamiltonian, the $\phi$ contribution is 
\be
\omega_\phi\sum_{i,j=1}^N\frac{a_{ij}^\phi a^{\dagger\phi}_{ji}+a^{\dagger\phi}_{ij}a^\phi_{ji}}{2}\rightarrow 0\;,
\ee
at least if we have $\omega_\phi N^2\rightarrow 0$ (for the $N^2$ degrees of freedom). Thus the full toy model Hamiltonian is actually 
\bea
\frac{\lambda}{N}H&=& \sum_{I=1}^ 3\sum_{i,j=1}^ N(2\omega)\frac{a^I_{ij}a^ {\dagger I }_{ji}
+a^ {\dagger I}_{ij}a^ I_{ji}}{2}\cr
&&+\frac{1}{2\sqrt{2}}\sum_{I,J,K=1}^ 3\sum_{i,j,k,l=1}^ N\epsilon_{IJK}(a^ I_{ij}
+a^ {\dagger I}_{ij})(a^ J_{jk}+a^ {\dagger J}_{jk})
(a^ K_{kl}+a^ {\dagger K}_{kl})\times\cr
&&\times\frac{i}{2\omega\sqrt{2\omega}} 
\left(2\omega\delta_{li}+\frac{\tilde g}{\sqrt{2\omega_\phi}}(a^\phi_{li}+a^{\dagger\phi}_{li})\right)\cr
&=&H_0+gH_{\rm int}\,.    \label{Hfulltoy}
\eea
We immediately observe that, in order for the limit
$\omega_\phi \rightarrow 0$
to remain well defined, the combination
$\tilde g/\sqrt{\omega_\phi}$
must be held fixed, even if it becomes parametrically large.
We also note that, relative to the coupling convention $g$ introduced in the previous subsection, the corresponding coupling in Maldacena’s notation, which we adopt here, is purely imaginary. This is in fact necessary for consistency. Taking the Hermitian adjoint of the cubic term involving the three $X$ fields produces an additional minus sign, which is precisely cancelled by the factor of $i$, ensuring that the Hamiltonian remains Hermitian.

\noindent
The same mechanism continues to hold once the field $\phi$ is inserted inside the trace. The adjoint operation again generates a minus sign, which is compensated by the explicit factor of $i$, so that the full interaction term remains real, as required. We define now the full vacuum state (which can be taken as the reference state) as 
\be
a^I_{ij}|0\rangle=0\;,\;\; a^\phi_{ij}=0\;,\;\; \forall I, i, j.
\ee
At spin $j=N$ (integer) and lower, we have the tensor product of highest weight spin 1 states
\be
(|J=1,-1\rangle\otimes ...\otimes |J=1,-1\rangle)_{k_1l_1...k_Nl_N}
=\frac{1}{\sqrt{N!}}\left(\frac{a^{\dagger 2}+ia^{\dagger 1}}{\sqrt{2}}\right)_{k_1l_1}...
\left(\frac{a^{\dagger 2}+ia^{\dagger 1}}{\sqrt{2}}\right)_{k_Nl_N}|0\rangle.
\ee
There are $3^N$ states in the tensor product of the spin 1 representations, 
decomposing, as usual, into a spin $N$ representation
plus many lower spin representations, making up the $3^N$ total number of states.
However, since we are interested in gauge invariant states only, we must consider only the single trace, 
or multi-trace states. 
The simplest state, the single trace state, is 
\be
|j=N,-N\rangle^{\rm single\;trace}=\frac{1}{\sqrt{N!}}\Tr\left[\left(\frac{a^{\dagger 2}
+ia^{\dagger 1}}{\sqrt{2}}\right)^N\right]|0\rangle.
\ee
As for the $N=2$ case, because of the cyclicity of the trace, and the fact that $a^\dagger$'s 
commute, this state is totally 
symmetric. But, in order to be naturally of spin $N$, we should also take out the traces; yet that will 
only impact the $m=0$ mode. We then need to act with $J_+$ on the above state, in order to get the states of other $m$. 
Again, because of cyclicity of the trace, plus the fact that $a^\dagger$'s all commute, 
so we can put a given combination in a 
fixed position inside the trace, we get the totally symmetric representation. After acting with 
$N$ $J_+$'s, we get the state of $m=0$, 
\be
|j=N,0\rangle^{\rm single\; trace}=\frac{(-1)^{N/2}}{\sqrt{N!}}\Tr[(a^{\dagger 3})^N+{\rm more}]|0\rangle\;,
\ee
where ``more" refers to things were, for instance $(a^{\dagger 3})$ is replaced by 
$(a^{\dagger 2}+ia^{\dagger 1})(a^{\dagger 2}-i
a^{\dagger 1})$. After acting with $N$ more $J_+$'s, we get the state of $m=N$, 
\be
|j=N,+N\rangle^{\rm single\; trace}
=\frac{1}{\sqrt{N!}}\Tr\left[\left(\frac{a^{\dagger 2}-ia^{\dagger 1}}{\sqrt{2}}\right)^N\right]|0\rangle.
\ee

%{\em I think the statement below is not true, and it only affects the $m=0$ mode, but I am not sure. 

%But, as before, we need to take out all the traces in order to get to the spin $N$ representation. For example, 
%the single trace inside 
%the $|j=N,-N\rangle^{\rm single\;trace}$ state is a generalization of the $N=2$ case, namely
%\be
%\Tr\left\{\left(\frac{a^{\dagger 2}+ia^{\dagger 1}}{\sqrt{2}}\right)^{N-2}
%\frac{1}{3}\left[\frac{1}{2}(a^{\dagger 2}+ia^{\dagger 1})^2-(a^{\dagger 3})^2
%+\frac{1}{2}(a^{\dagger 2}-ia^{\dagger 1})^2\right]\right\}
%|0\rangle\;,
%\ee}

\subsection{On the growth of $a_n,b_n$ in the perturbative region}

We are able to estimate the growth of the early Lanczos coefficients by using the scheme outlined in \cite{Haque:2022ncl}.  We first do this for the simplified version (\ref{Hsimplified}).  
By defining the usual creation and annihilation operators, the above Hamiltonian takes the form
\begin{eqnarray}
g H_{int} & = &  L_{+} + l_{+} + l_{-} 
+ L_{-}   \nonumber \\
L_{+} & = &  \frac{g}{2\sqrt{2}}\sum_{I,J,K=1}^3 \sum_{i,j,k=1}^ N\epsilon_{IJK} a^ {\dagger I}_{ij} a^ {\dagger J}_{jk}
a^ {\dagger K}_{ki} \nonumber \\
& = & \frac{3 g}{2 \sqrt{2}} \left(  Tr\left( a^{\dag 1} a^{\dag 2} a^{\dag 3}   \right)      -    Tr\left( a^{\dag 3} a^{\dag 2} a^{\dag 1}   \right)  \right)\nonumber \\
l_{+} & = &  \frac{g}{2\sqrt{2}}\sum_{I,J,K=1}^3 \sum_{i,j,k=1}^ N\epsilon_{IJK}( a^ {\dag I}_{ij} a^ {\dag J}_{jk}
a^ {K}_{ki}  + a^ {\dag I}_{ij} a^ {J}_{jk}
a^ {\dag K}_{ki}  + a^ { I}_{ij} a^ {\dag J}_{jk}
a^ {\dag K}_{ki} )  \nonumber \\
& = & \frac{3g}{2 \sqrt{2}} Tr\left( \left[  a^{\dag 1},  a^{\dag 2} \right] a^{3} + \left[ a^{\dag 3},  a^{\dag 3} \right] a^{ 2}  +  \left[ a^{\dag 2},  a^{\dag 3} \right] a^{ 1}  \right)   \nonumber \\
l_{-} & = & l_{+}^\dag   \nonumber \\
L_{-} & = & L_{+}^\dag \nonumber 
\end{eqnarray}
As reference state we choose the vacuum.  We can readily identify a family of states that satisfy the conditions (\ref{AppCond1}) and (\ref{AppCond2}).   This is the family of states:
$$   | K_n^{(0)} ) = L_{+}^n |0\rangle = \left( \frac{3g}{2 \sqrt{2}} \right)^n \sum_{m=0}^n (-1)^m \Tr(a^{\dag 1} a^{\dag 2} a^{\dag 3})^{n-m} \Tr(a^{\dag 3} a^{\dag 2} a^{\dag 1})^m |0\rangle .$$
These are unnormalized.  Indeed, the Lanczos coefficients are precisely related to a ratio of the 
normalisation factors 
$$ b_{n+1}^{(0)} =   \sqrt{ \frac{ \langle 0 | L_{-}^{n+1} L_{+}^{n+1}   
| 0 \rangle    }{ \langle 0 | L_{-}^{n} L_{+}^{n}   | 0 \rangle}           }    .$$
The operator $L_{+}$ consists of two terms.  First, let us just consider the family of states
$$ | A_n) =   \Tr( a^{\dag 1} a^{\dag 2} a^{\dag 3}   )^n |0\rangle  .$$
To compute the overlap $(A_n)$ we need to compute all possible Wick contractions. 
This gives a sum over permutations.  This is similar to the computations performed in 
\cite{Corley:2001zk}.   We find
\begin{eqnarray}
( A_n |    A_n   )  
& = & \sum_{\sigma \in S_n} \sum_{\tau \in S_n} \sum_{\rho \in S_n} \sum_{i_1,\cdots, i_n} 
\sum_{j_1, \cdots, j_n} \sum_{k_1, \cdots, k_n} \sum_{i'_1, \cdots, i'_n} \sum_{j'_1, \cdots, j'_n}  \sum_{k'_1, 
\cdots, k'_n} \delta^{i_m}_{i'_{\sigma(m)}} \cr
&&\delta_{j_m}^{j'_{\sigma(m)}}   
%\ \ \ \ \ \ \ \ \ \ \ \ \ \ \ \ \ \ \ \ \ \ \ \ \ \ \ \ \ \ \ \ \ \ \ \ \ \ \ \ \ \ \ \ \ \ \ \ \ \ \ \ \ \ \ \ \ \ \ \ \ \ \ \ \ \ \ \ \ \ \ \ \ \ \ \ \ \ \ \  
\delta^{j_m}_{j'_{\tau(m)}} \delta_{k_m}^{k'_{\tau(m)}} \delta^{k_m}_{k'_{\rho(m)}} 
\delta_{i_m}^{i'_{\rho(m)}} \nonumber \\
& = & \sum_{\sigma \in S_n} \sum_{\tau \in S_n} \sum_{\rho \in S_n} \sum_{i'_1, \cdots, i'_n} 
\sum_{j'_1, \cdots, j'_n}  \sum_{k'_1, \cdots, k'_n} \delta^{i'_{\rho(m)}}_{i'_{\sigma(m)}}\delta^{j'_{\sigma(m)}}
_{j'_{\tau(m)}} \delta^{k'_{\tau(m)}}_{k'_{\rho(m)}}     \nonumber \\
& = &  n! \sum_{\tau \in S_n} \sum_{\rho \in S_n} \sum_{i_1, i_2, \cdots, i_n}\sum_{j_1, j_2, 
\cdots, j_n} \sum_{k_1, k_2, \cdots, k_n} \delta^{k_m}_{k_{\tau(m) } } \delta^{i_{\rho(m)}}_{i_m}   
\delta^{j_{\tau(m)}}_{j_\rho(m)}           \nonumber \\
& = &  n! \sum_{\tau \in S_n} \sum_{\rho \in S_n}  \sum_{j_1, j_2, \cdots, j_n} N^{C(\tau) + 
C(\rho)}  \delta^{j_{\tau(m)}}_{j_\rho(m)}     \nonumber \\
& = &  n! \sum_{\tau \in S_n} \sum_{\rho \in S_n}  N^{C(\tau) + C(\rho) + C(\tau \rho^{-1})} \;,
\label{AnAn}  
\end{eqnarray}
where $C(\sigma)$ is the number of cycles in the permutation $\sigma$. 
This is an exact result (though we still have to compute the sum over permutations).  When 
considering the large $N$ limit, we get differing behavior depending on how $n$ scales with $N$, i.e., 
which regime of complex states are we interested in.  When $n$ is order $1$ then the dominant term 
in the permutation summation is when $\tau = \sigma = I$, which gives
$$ 
( A_n |    A_n   )  \sim  n! N^{3n}\left( 1 + O\left(\frac{1}{N}\right)    \right).
$$
In this limit the different traces also don't mix so that, in fact, we have
$$ \langle 0 | L_{-}^n  L_{+}^n | 0 \rangle \sim n! \left( \frac{3g}{\sqrt{2}}\right)^{2n} N^{3n}\left( 1 + O\left(\frac{1}{N}\right)    \right) \;,  $$
and the approximate Lanczos coefficients grow as $b_n^{(0)} \sim \frac{3 g}{\sqrt{2}} N^{\frac{3}{2}}\sqrt{n}$.  We also find that $a_n^{(0)} \sim 6 \omega n$.  This gives the perturbative behaviour that we were looking for. In contrast, when $n$ is much larger than $N$ (e.g. $n$ scales as $N^2$), then the dominant 
contributions above should come from $\tau$ and $\sigma$ being $n$-cycles\footnote{The result for two matrices is quite insightful to consider in this regard.  See appendix \ref{SingleCylceSum} for details.  }.  In this regime we also cannot ignore the mixing of different trace structures.  For this reason these results should be considered to be vailid only for small $n$.  \\ \\
%\textbf{To do:} What is the appropriate expression for this?
Next, we consider the Hamiltonian (\ref{Hfulltoy}) with the vacuum chosen as reference state. For this Hamiltonian we can again find an approximate Krylov basis satisfying the necessary conditions (\ref{AppCond1}) and (\ref{AppCond2}).   These are a simple adjustment of the approximate Krylov basis for the simplified version,
$$  |B_n) = \Tr(a^{\dag \phi} a^{\dag 1} a^{\dag 2} a^{\dag 3})^n |0\rangle .$$
For the overlap we find
\begin{eqnarray}
(B_n| B_n) & = & \sum_{\sigma\in S_n}\sum_{\tau \in S_n} \sum_{\mu\in S_n} \sum_{\nu \in S_n} 
\sum_{i_1, \cdots, i_n} \sum_{j_1, \cdots, j_n} \sum_{k_1, \cdots, k_n} \sum_{l_1, \cdots, l_n} 
\delta^{i_{\sigma(m)}}_{i_{\tau(m)}} \delta^{j_{\tau(m)}}_{j_{\mu(m)}} \delta^{k_{\mu(m)}}_{k_{\nu(m)}} 
\delta^{l_{\nu(m)}}_{l_{\sigma(m)}}    \nonumber \\
& = & n! \sum_{\tau \in S_n} \sum_{\mu\in S_n} \sum_{\nu \in S_n} \sum_{i_1, \cdots, i_n} \sum_{j_1, \cdots, 
j_n} \sum_{k_1, \cdots, k_n} \sum_{l_1, \cdots, l_n}  \delta^{i_{m}}_{i_{\tau(m)}} \delta^{j_{\tau(m)}}
_{j_{\mu(m)}} \delta^{k_{\mu(m)}}_{k_{\nu(m)}} \delta^{l_{\nu(m)}}_{l_{m}}    \nonumber  \\ 
& = & n! \sum_{\tau \in S_n} \sum_{\mu\in S_n} \sum_{\nu \in S_n} N^{ C(\tau) + C(\nu) + C(\tau \mu^{-1}) + 
C(\nu \mu^{-1})    }.   \nonumber 
\end{eqnarray}
When $N$ is large and $n$ is order $1$ then the dominant contribution again comes from all three cycles equation to the identity.  This gives
$$ (B_n|B_n) \sim n! N^{4 n}(1 + O(\frac{1}{N}))$$
and again gives the approximate growth as $b_{n}^{(0)} \sim \sqrt{n}$.  \\ \\
A natural question to ask is how accurate these approximate Lanczos coefficients are expected to be. 
Since the early time growth of complexity is governed by the first few Lanczos coefficients\footnote{The $t^{n}$ contribution is made up of combinations of the first $n$ Lanczos coefficients.} we need to estimate how accurate the approximate Krylov basis is.  One way to estimate this is to consider the ratio
$$ \frac{\langle 0 |H^{2n} |0\rangle}{\langle 0 | (L_{-})^n(L_{+})^n |0\rangle     }$$
and see whether it is close to 1.  If so, then the approximate Lanczos coefficients provide an accurate description (i.e. $\cos\theta_n \sim 1$).  The state
$$H^n |0\rangle$$ 
consists of different combinations of the operators $L_{+}, L_{-}, l_{+}, l_{-}$ and $H_0$.  These may be organised in terms of the number of creation operators (i.e. in terms of the eigenvectors of $H_0$) appearing inside the traces.  Each action of $L_{+}$ increases this number by $3$, $l_{+}$ by $1$ and $H_0$ by $0$.  At small $n$ (where different trace structures do not mix) we find that the overlap of 
$$ |\psi_{n; k, l}) \equiv L_{+}^{n-k-l}l_{+}^{k}H_{0}^l |0\rangle$$ 
scales as $g^{n-l}  N^{3n - 2k - 3l}$ which is the dominant contribution at large $N$.  It thus follows that 
$$ \frac{\langle 0 |H^{2n} |0\rangle}{\langle 0 | (L_{-})^n(L_{+})^n |0\rangle     } = \frac{c_{n; 0, 0} g^n N^{3n} + \sum_{k, l } c_{n; k, l} g^{n-l}  N^{3n - 2k - 3l} }{c_{n; 0, 0} g^n N^{3n} } = 1 + \sum_{k, l } \frac{c_{n; k, l}}{c_{n; 0, 0}} (N^3 g)^{-l}  N^{- 2k}   $$
For small $n$ and $g >> \frac{1}{N^2}$ we thus find that the approximate Krylov basis provides an accurate description of the dynamics.

\section{Late time behaviour and instantons}

%\textbf{To do:} Probably need to consider asymptotic formulas for large partitions to get a proper handle on this.  

\subsection{Toy model states relevant for late times}

Next, we consider some arguments about the onset of the later time behaviour and the plateau. The first observation is that at late times, $e^{i\hat H t}$ acting on a state means the multiple (large $j$ times)
commutator $[H,[H,[H...]]]$ acting on a state which, as we constructed, is obtained from the action of
$a^\dagger$'s on $|0\rangle$. Since $H$ contains $(a+a^\dagger)^3$, this will contain terms that 
increase the spin of the 
state. Since moreover the hamiltonian is $SO(3)$ invariant, that means that the late time behaviour 
for the state is defined by a large spin $j$ (representation of $SO(3)$). We also have the action of 
$a_\phi^\dagger$ (radiation) on the states, so the relevant states should be 
\be
``\,(a^{\dagger \phi})^N|j=N, -N\rangle^{\rm single\; trace}\,"\equiv 
\frac{1}{\sqrt{N!}}\Tr\left[(a^{\dagger \phi})^N\left(\frac{a^{\dagger 2}
+ia^{\dagger 1}}{\sqrt{2}}\right)^N\right]|0\rangle
\ee
with the traces properly subtracted, and all of its descendents in the spin $N$ representation.
This constitutes a guess for the maximum complexity states at large times (and large $N$). 

\subsubsection{Complexity at large $j$}

We now present a heuristic picture for the late-time behaviour of the complexity based on the large-spin structure of the relevant states. Let us begin with the free part of the toy-model Hamiltonian. Omitting the vacuum zero-point contribution,
$E_0=\sum \frac{\omega}{2}$,
the excitation energy is
\be
E_{{n}}-E_0
=
\sum_{i,j=1}^N
\sum_{I=1}^3
n_{ij}^I ,\omega
\equiv
\tilde N,\omega.
\ee
As discussed in the previous subsection, the total number of creation operators $a^\dagger$ determines the total spin quantum number $j$ of the corresponding $SU(2)\simeq SO(3)$ representation. Explicitly,
\be
j
=
\tilde N
=
\sum_{i,j=1}^N
\sum_{I=1}^3
n_{ij}^I\,.
\ee
The excitation energy may therefore be written simply as
\be
E_{{n}}-E_0
=
j,\omega.
\ee
Both the free Hamiltonian and the cubic interaction term commute with the generators $J^i$ of $SU(2)$, and the system is therefore $SU(2)$ invariant. In general, the states may thus be labelled by
$|Ejm\rangle$,
where $m$ labels the states within the spin-$j$ representation of dimension $2j+1$, while $E$ is, a priori, an independent energy label.\\

\noindent
For the free theory, however, the relation
$E-E_0=j\omega$
identifies the energy directly with the spin quantum number. The states may therefore be labelled simply by $|jm\rangle$. In this picture, the free time evolution corresponds to motion within a fixed finite-dimensional $SU(2)$ representation of dimension $2j+1$. The role of the interaction is then to populate progressively larger representations as time evolves, effectively driving the dynamics toward states with increasing total spin $j$. \\

\noindent
There is, however, additional structure beyond the global spin labels. The states are also characterized by the individual occupation numbers
$\{n_{ij}^I\}$, and, if one restricts to gauge-invariant sectors, potentially also by the number of traces appearing in the corresponding operators. More precisely, the relevant states should therefore be viewed schematically as
$|jm;\{n_{ij}^I\}\rangle_{\rm n\text{-}trace}$. Now consider the large-$N$ regime with
\be
j=\tilde N \ll N\rightarrow\infty,
\qquad
{\rm or\ perhaps\ more\ generally}
\qquad
j=\tilde N < N\rightarrow\infty.
\ee
In this limit, states initially lying in the single-trace sector of a fixed spin-$j$ representation are expected to remain approximately within that sector during the evolution. The reason is that transitions to multi-trace sectors are suppressed by powers of $1/N$, while energy conservation fixes the total occupation number
$\tilde N = \sum_{ijI} n_{ij}^I$,
even though the individual occupation numbers $n_{ij}^I$ may still redistribute dynamically.
The number of possibilities of fixed $j=\tilde N$, giving the dimension of the Hilbert space 
of the evolving system corresponds to: 

{\em \# of partitions of $j=\tilde N$ into $n_{ij}^I= 3N^2$ numbers, 
times $(2\tilde N+1)$}.\\

\noindent
In the regime $\tilde N<N$, the finite-$N$ cutoff on the occupation numbers is not yet important. The dominant counting problem is therefore the number of ways of distributing the total excitation number $\tilde N$ among oscillator modes. To leading order at large $\tilde N$, this is controlled by the partition number
\be
p(\tilde N)
\simeq
\frac{1}{4\tilde N\sqrt{3}}
\exp\left(
\pi\sqrt{\frac{2\tilde N}{3}}
\right).
\ee
Including the degeneracy of the spin-$j=\tilde N$ representation gives an additional factor
$2\tilde N+1$,
so that the number of states available at fixed $\tilde N$ scales as
\be
{\cal N}(\tilde N)
\sim
(2\tilde N+1)p(\tilde N)
\simeq
\frac{1}{2\sqrt{3}}
\exp\left(
\pi\sqrt{\frac{2\tilde N}{3}}
\right),
\ee
up to subleading algebraic factors.\\

\noindent
There is an important caveat. If one restricts strictly to single-trace states, rather than allowing generic gauge-invariant states, the precise counting is more subtle. The single-trace constraint does not simply reduce to unrestricted integer partitions, and a more careful treatment would require counting cyclic words built from the available matrix oscillators at fixed total excitation number. Nevertheless, the estimate above gives the natural scale of the Hilbert-space dimension available before finite-$N$ trace relations become important. In the free theory, time evolution within a fixed spin-$j$ sector is quasi-periodic, as in the motion of a particle on an $SU(2)$ representation of dimension $2j+1$. For example, for a Hamiltonian of the form
$H=\alpha(J_+ + J_-)$,
one obtains an oscillatory complexity of the schematic form
$C(t)\sim 2j\sin^2(\alpha t)$.
Thus the free theory explores only a finite representation and does not generate irreversible spreading through the full space of states. The cubic interaction term, schematically $\epsilon XXX$, changes this picture. Although the full Hamiltonian remains $SO(3)$ invariant, the interaction mixes the occupation-number configurations ${n_{ij}^I}$ within a fixed total energy shell and, more generally, allows transitions between increasingly large effective spin sectors as the system evolves. In the large-$N$ regime, this mixing is expected to be sufficiently complicated to produce chaotic motion on the accessible Hilbert space. The late-time Krylov dynamics should then be controlled not by coherent motion in a single $SU(2)$ multiplet, but by the effective dimension of the energy shell explored by the interacting dynamics.\\

\noindent
A more robust way to state the argument is therefore the following. The Krylov complexity cannot grow beyond the dimension of the dynamically accessible Hilbert space. If the interaction efficiently scrambles the states at fixed excitation number, then the plateau value should scale parametrically as
\be
C_{\rm plateau}
\sim
{\cal N}(\tilde N)
\sim
\exp\left(
\pi\sqrt{\frac{2\tilde N}{3}}
\right),
\ee
up to algebraic prefactors.
Equivalently, if the relevant excitation number reached by time $t$ grows approximately linearly,
$\tilde N(t)\sim t$,
then one obtains the stretched-exponential estimate
\be
C(t)
\sim
\exp\left(
\pi\sqrt{\frac{2t}{3}}
\right),
\ee
before saturation. If instead the growth of $\tilde N(t)$ is diffusive, $\tilde N(t)\sim \sqrt t$, the corresponding growth would be slower,
$C(t)\sim \exp\left({\rm constant}\times t^{1/4}\right)$.
Thus the precise time dependence depends on the transport law in excitation space, but the plateau scale is fixed by the density of accessible states. In the simple ballistic estimate where the system spreads across the spin representation over a time
\be
t_{\rm max}\sim 2\tilde N,
\ee
one obtains
\be
C_{\rm plateau}
\sim
C(t_{\rm max})
\sim
\frac{1}{4\sqrt{3}}
\exp\left(
\pi\sqrt{\frac{t_{\rm max}}{3}}
\right),
\ee
again up to algebraic prefactors. This provides a natural explanation for the emergence of a plateau and for the stretched-exponential scaling suggested by the state-counting argument. While this reasoning is not a derivation, it gives a physically sharper interpretation of the late-time behaviour; the plateau is controlled by the entropy of the accessible large-spin Hilbert space, while the approach to the plateau depends on how efficiently the interaction explores that space.

%{\em But there are many hand-waving arguments here, and one missing item 
%(how to calculate the number of single trace states). }

\subsection{Plateau value from quantum mechanical instantons}

Here we start by reviewing our previous work in \cite{Beetar:2025erl}, 
for the general plateau value for the complexity of 
chaotic quantum mechanical systems, and then we apply it to the case in this paper. At large times, the Krylov complexity $C(t)$ for chaotic systems is supposed to obtain go to a 
plateau value, see for instance \cite{Balasubramanian:2022tpr}. But the Kyrlov complexity is an 
infinite sum involving the coefficients $\psi_n(t)$ of the expansion of the wave function $|\psi(t)\rangle$ 
into the Krylov basis $|n\rangle$, 
\be
\psi_n(t)=\langle n|\psi(t)\rangle=\langle n|e^{-\frac{i}{\hbar} Ht}|\psi\rangle.
\ee
For a quantum mechanical system with $N$ degrees of freedom, with coordinates $\{x_i\}_{i=1,N}$, 
introducing the identity over this Hilbert space with basis $|\{x_i\}\rangle$, we obtain 
\be
\psi_n(t)=\left(\prod_{i}\prod_{j}\int dx_{i}\int dx'_{j}\right)
\; K_n^*(\{x_{i}\})\psi(\{x'_{j}\})\langle \{x_{i}\}|e^{-\frac{i}{\hbar}Ht}|\{x'_{j}\}\rangle\;.
\ee
The resulting matrix element can be written as a path integral in the usual way, 
\be
\langle \{x_i(t)\}|\{x'_j(0)\}\rangle=\langle\{x_i\}|e^{-iHt}|\{x'_j\}\rangle=
\int _{\{x_i(t)\};\{x'_j(0)\}}{\cal D} \{x_m(t')\}e^{iS[\{x_m(t')\}]}\;,
\ee
which leads to the Krylov complexity formula
\bea
K(t)&=&\sum_n n\left| \left(\prod_{i=1}^N\prod_{j=1}^N\int dx_{i}\int dx'_{j}\right)
\; K_n^*(\{x_{i}\})\psi(\{x'_{j}\})\right.\times\cr
&&\left.\times\int _{\{x_i(t)\};\{x'_j(0)\}}{\cal D} \{x_m(t')\}e^{iS[\{x_m(t')\}]}\right|^2.
\eea
Without the integral over the endpoints $x_i, x_f$, the transition amplitude given by the path integral 
would peak over the classical action, {\em if there is a classical path between them}, 
$\langle x(t)|x'(0)\rangle \simeq e^{iS_{\rm cl}[x(t);x'(0)]}$. But if there is no classical path, the 
transition amplitude is given instead 
by a quantum mechanical instanton solution, which is to say by a classical solution in 
Euclidean time $t_E=-it$ (which is an imaginary solution in terms of real time), for motion in the 
inverted potential $V_E=-V$. The instanton is standard and
best understood in the case of $T_E=2t_E\rightarrow\infty$, 
between two maxima of $V_E$, or minima of $V$, in one of them at $t_E\rightarrow -\infty$ and another
at $t_E\rightarrow+\infty$.\\ 

\noindent
Since we want to consider the $t\rightarrow\infty$ limit 
(so $it=T_E\rightarrow\infty$), yet integrating over 
$x(t)=x_f$ and $x(0)=x_i$, and {\em in the case of chaotic systems}. Then $e^{iS}$ will be a wildly varying 
phase, averaging out to zero under the integration over the {\em classical, or real time, paths} and their 
endpoints, leaving only the complex (imaginary time) paths of instantons, indicating quantum transitions. 
The relevant chaotic systems must then have at least 2 local minima for $V(x)$, and integrating over $x_i$ 
and $x_f$ and $t$ will lead to a result sharply peaked on the minimal $S_E$ (saddle point approximation), 
which is for the {\em standard instanton  between two minima $x_1,x_2$ of $V(x)$},
\be
\int dx_i \int dx_f e^{-S_E}\simeq \sum_{x_1,x_2={\rm minima}}e^{-S_E(x_i=x_1,x_f=x_2)}.
\ee
This gives a plateau value for $\psi_n(t\rightarrow\infty)$, 
\bea
\psi_n(t\rightarrow\infty) &\simeq& \sum_{\{x_1\},\{x_2\}={\rm minima}}K_n^*(\{x_2\}) \psi(\{x_1\})
\langle \{x_2(t\rightarrow\infty)\}|\{x_1(0)\}\rangle\cr
&\simeq& \sum_{\{x_1\},\{x_2\}={\rm minima}}K_n^*(\{x_2\}) \psi(\{x_1\}) e^{-S_E({\rm instanton}; x_1,x_2)}\;,
\eea
and so a plateau value for the Krylov complexity,
\bea
K(t\rightarrow\infty)
&=&\sum_n n|\psi_n(t\rightarrow\infty)|^2\cr
&\simeq& \sum_n n \left| \sum_{\{x_1\},\{x_2\}
={\rm minima}}K_n^*(\{x_2\}) \psi(\{x_1\}) e^{-
S_E({\rm instanton}; x_1,x_2)}\right|^2\;,
\eea
which is topological (in the restricted sense that it depends only on the minima of the potential). If the 
potential has {\em only} two minima, we get
\be
K(t\rightarrow\infty)
\simeq e^{-S_E({\rm instanton})}\sum_n n \left|K_n^*(\{x_2\})\psi(\{x_1\})+K_n^*(\{x_1\})
\psi(\{x_2\})\right|^2.\label{compplateau}
\ee
Note that we are actually interested mostly in the Krylov entropy, which has a log of $|\psi_n|^2$, 
harder to deal with, but 
for it we can use the following argument. We can define the density matrix operator
\be
\hat \rho= \hat \rho_A\equiv \sum_n |\psi_n|^2|n\rangle\langle n|\;,
\ee
and from it the $N$-th order partition function
\be
Z_N=\Tr\hat \rho^N=\sum_n \langle n|\left(\sum_m |m\rangle\langle m| (|\psi_m|^2)^N\right)|n\rangle
=\sum_n |\psi_n|^{2N}\;,
\ee
where we have used the orthonormality of the Krylov basis $|n\rangle$. But then the Krylov entropy is obtained from it from a limit and a derivative, 
\be
S_K=-\sum_n |\psi_n|^2\log |\psi_n|^2=\lim _{N\rightarrow 1}(1-N\d_N)\log \sum_n |\psi_n|^{2N}
=\lim_{N\rightarrow 1}(1-N\d_N)\log Z_N\;,
\ee
just as in the case of the entanglement entropy.\\

\noindent
Applying these ideas to matrix models with $U(N)$ symmetry introduces an important additional subtlety namely the treatment of gauge invariance.
In both the BFSS and BMN matrix models, the dynamics is originally formulated in terms of covariant derivatives
\be
D_t=\partial_t+A_0,
\ee
acting on the matrix fields $X^i$. The theory is therefore invariant under time-dependent $U(N)$ gauge transformations,
\be
X^i(t)\rightarrow U^\dagger(t)X^i(t)U(t),
\ee
together with
\be
A_0(t)\rightarrow
U^\dagger(t)A_0(t)U(t)
+
U^\dagger(t)\partial_t U(t).
\ee
Since the theory is one-dimensional, the gauge field has only the temporal component $A_0$. One usually fixes the gauge by setting
$A_0=0$,
thereby eliminating the gauge field entirely from the dynamical description. However, this does not remove all gauge redundancy. After fixing $A_0=0$, one is still left with residual time-independent $U(N)$ transformations,
$X^i \rightarrow U^\dagger X^i U$,
which act as global symmetries on the matrix variables. These residual transformations must still be accounted for in the path integral.\\

\noindent
In addition, varying the action with respect to $A_0$ before gauge fixing produces the Gauss-law constraint, which must continue to be imposed on physical states. More fundamentally, the path integral still contains the integration
\begin{eqnarray}
    \int_{U(N)} dU,
\end{eqnarray}
corresponding to the residual constant unitary transformations acting on the matrices $X_{ij}(t)$. Properly implementing this remaining gauge identification is therefore an essential part of defining the matrix-model dynamics.\\

\noindent
A different approach to this issue was proposed in \cite{Fliss:2025kzi} for matrix models, including the bosonic mini-BMN model considered here (though in that work the quartic scalar interaction was retained). Rather than working directly with explicitly gauge-invariant operators, the authors introduced so-called relational operators. These are defined by dressing the matrix variables with a group element $U\in U(N)$,
\be
\hat X^a_{{\bf 1},ij}
\equiv
(\hat U^\dagger X^a \hat U)_{ij},
\ee
where
\be
\hat U
=
\int_{U(N)} dU \, U \hat P_U,
\qquad
\hat P_U
=
|U\rangle\langle U|.
\ee
The essential idea is that gauge invariance is implemented through an averaging over the group orbit, rather than by restricting attention solely to explicitly gauge-invariant combinations of the matrix variables. In \cite{Fliss:2025kzi}, the authors were interested in transitions between equivalence classes associated with subgroups such as $U(M)$ and $U(N-M)$, while treating the full $U(N)$ as the gauge redundancy to be integrated over. They then further restricted the integration domain in a manner consistent with the residual subgroup symmetries.\\

\noindent
Motivated by this construction, it is natural in our case to consider similarly averaged operators such as
\be
\hat\rho
=
\int_{U(N)} dU \,\hat\rho_U,
\qquad
\hat K
=
\int_{U(N)} dU \,\hat K_U,
\ee
where
\be
\hat K
=
\sum_n n\,|n\rangle\langle n|
\ee
is the Krylov complexity operator. The operators $\hat\rho_U$ and $\hat K_U$ are not themselves gauge invariant; rather, they should be understood as relational operators in the sense above. The Krylov complexity is then given by the expectation value
\be
K(t)
=
\langle\psi(t)|\hat K|\psi(t)\rangle.
\ee
Following the reasoning of \cite{Fliss:2025kzi}, the remaining integration over $U(N)$ is expected to be dominated by a saddle-point configuration in the large-$N$ limit. Under this assumption, the group averaging does not qualitatively modify the leading large-$N$ dynamics, and one may treat the gauge integration semiclassically.
Instead, we must just consider the integration over variables $X_{ij}(t)$ as in the generic case described 
previously. Then 
\bea
\psi_n(t)&=&\langle n|\psi(t)\rangle=\langle n|e^{-\frac{i}{\hbar} Ht}|\psi\rangle\cr
&=&
\left(\prod_{ij}\prod_{kl}\int dX_{ij}\int dX'_{kl}\right)
\; K_n^*(\{X_{ij}\})\psi(\{X'_{kl}\})\langle \{X_{ij}\}|e^{-\frac{i}{\hbar}Ht}|\{X'_{kl}\}\rangle,
\eea
and use the path integral for the matrix element (transition amplitude), 
\be
\langle X_{ij}(t)|X'_{kl}(0)\rangle=
\langle X_{ij}|e^{-\frac{i}{\hbar}Ht}|X'_{kl}\rangle=\int_{X_{ij}(t);X'_{kl}(0)}{\cal D}X_{mn}(t')e^{iS[X_{mn}(t')]}
\delta_{ij}^{kl}.
\ee
Then we have for the transition amplitude
\be
\langle X_{ij}|e^{-\frac{i}{\hbar}Ht}|X'_{kl}\rangle\simeq \sum_p \int dU e^{-S_{\rm inst}^{(p)}[X_{ij},X'_{kl};U]}
\delta_{ij}^{kl}\;,
\ee
where $p$ are Euclidean instanton saddle points of the path integral (since the matrix model must be 
chaotic, as it represents a black hole). The integral over $U$ reduces to a saddle point as well, and the 
rest proceeds as in the general case, leading to a Krylov complexity plateau, and thus also to a 
plateau for the Krylov entropy.

\subsection{Scaling behaviour of Krylov entropy}

For the BMN D0-brane black hole in 9+1 dimensions, the expected entropy formula is \cite{Maldacena:2023acv} 
\be
S\propto N^2 \left(\frac{T}{\lambda^{1/3}}\right)^{9/5}\;,\label{S10d}
\ee
a very nontrivial power. However, we are really considering the mini-BMN model, for which we have found that the entropy formula replacing the 10-dimensional formula (see, for example, \ref{S10d}) was (\ref{BMNent}), i.e., 
\be
S\propto N^2\left(\frac{T}{\lambda^{1/3}}\right)^{9/20}.
\ee
Note that the conditions on the BMN D0-brane model (\ref{BMNcond}) can be rewritten as 
\be
\frac{\omega}{T}\ll 1\;,\;\; \frac{T}{\lambda^{1/3}}\ll 1\;,\;\;\;
N\left(\frac{T}{\lambda^{1/3}}\right)^{9/5}\sim \frac{S}{N}\gg 1\;,
\ee
but maybe one needs to replace the last one, in the case of our model, with 
\be
\frac{S}{N}\sim N\left(\frac{T}{\lambda^{1/3}}\right)^{9/20}\gg 1
%=\left(\frac{T}{N^{10/9}\lambda^{1/3}}\right)^{9/20}\gg 1\;,
\ee
or (we still need to have $\lambda/T^3\gg 1$, and)
\be
T\gg \lambda^{1/3}N^{20/9}\Rightarrow \frac{\lambda}{T^3}\ll N^{20/3}.
\ee
This is, of course, a general argument about entropy, but in particular it should apply to Krylov entropy, 
which we can calculate from the Krylov basis decomposition.

\section{Conclusions}

In this paper we have investigated Krylov complexity and Krylov entropy in a simplified quantum mechanical model intended to capture essential aspects of a black hole interacting with its radiation. Our starting point was the BMN matrix model, argued in \cite{Maldacena:2023acv} to provide one of the simplest nonperturbative descriptions of near-extremal black holes in warped $AdS_2$. Motivated by the difficulty of analyzing the full BMN dynamics directly, we constructed a simplified mini-BMN-inspired toy model that retains the features most relevant for information-theoretic observables while remaining amenable to explicit perturbative and numerical analysis.\\

\noindent
A central theme of this work has been the interplay between early-time operator growth and late-time equilibration. At early times, perturbative and numerical calculations indicate the expected growth of Krylov complexity characteristic of strongly interacting quantum systems. At the same time, the associated Krylov entropy provides a natural information-theoretic probe of the spreading of amplitudes in Krylov space. While these observables are not identical to conventional notions of entanglement entropy, their behaviour nevertheless captures qualitative features expected of black-hole information dynamics.
The late-time regime is particularly interesting. In asymptotically AdS-like settings, radiation emitted by the black hole may effectively return from the boundary, leading to equilibration between the black hole and its radiation bath. Correspondingly, one expects information-theoretic observables to saturate rather than grow indefinitely. In the present model, this manifests itself through plateau behaviour in Krylov observables. Building on the path-integral formulation developed in \cite{Beetar:2025erl}, we argued that this saturation behaviour admits a semiclassical interpretation in terms of Euclidean instanton configurations governing the late-time dynamics. From this perspective, the plateau value is controlled by nonperturbative saddles of an effective Krylov action.\\

\noindent
One of the broader lessons of this analysis is that Krylov complexity appears to interpolate naturally between two physically distinct regimes; an early-time scrambling regime characterized by rapid operator growth, and a late-time equilibration regime controlled by finite entropy and nonperturbative effects. In this sense, Krylov observables may provide a useful bridge between microscopic quantum dynamics and emergent thermodynamic behaviour in gravitational systems. At a more technical level, the analysis also suggests that simplified matrix-model truncations can retain surprisingly robust information-theoretic features of the full theory. In particular, the mini-BMN-inspired truncation considered here appears sufficient to reproduce qualitative aspects of black-hole thermalization and saturation while avoiding many of the technical complications of the full BMN matrix model.\\

\noindent
There are several natural directions for future work:
\begin{itemize}
    \item A first priority is to extend the present analysis beyond the bosonic truncation and incorporate the full supersymmetric structure of the mini-BMN model. Supersymmetry may significantly modify the late-time structure of Krylov observables, particularly through cancellations between bosonic and fermionic sectors and through the existence of protected states.
    \item A second important direction would be to understand the role of the quartic commutator interaction more systematically. In the present work we argued that there exist restricted low-energy regimes in which the cubic radiation coupling dominates. Nevertheless, a more complete treatment should ultimately include the full strongly coupled matrix dynamics and determine quantitatively how the quartic interaction modifies operator growth and saturation. It would also be interesting to investigate the relation between Krylov entropy and more conventional entropic observables in gravitational systems. In particular, one would like to understand whether there exists a more direct connection between Krylov entropy and Page-curve physics, perhaps through the Thermo-Field Double construction or through a replica-wormhole-inspired formulations of Krylov dynamics.
    \item Another promising direction concerns the effective path-integral description itself. The present work provides further evidence that Krylov dynamics admits a semiclassical formulation in terms of an emergent effective geometry on Krylov space. It would be very interesting to understand whether this structure can be embedded directly into the original path integral of the underlying matrix model, or whether it defines an independent emergent semiclassical sector associated with operator growth along the lines of \cite{Murugan:2026yyu,Bhattacharyya:2026qef}.
\end{itemize}
\noindent
More generally, one may hope that Krylov observables provide a useful framework for studying quantum gravity beyond the specific BMN setting considered here. Possible extensions include SYK-like systems, matrix models dual to higher-dimensional black holes, open quantum systems coupled to baths, and time-dependent or evaporating black-hole geometries. It would also be interesting to investigate whether related plateau phenomena arise in systems with different holographic asymptotics or in models exhibiting confinement/deconfinement transitions. Finally, the present work reinforces the broader idea that complexity-theoretic observables may play a fundamental role in the microscopic description of black holes. While entanglement entropy captures static aspects of quantum correlations, Krylov complexity and Krylov entropy appear to probe dynamical aspects of information spreading and equilibration that are not directly visible through conventional thermodynamic observables alone. Understanding this relationship more deeply may ultimately shed light on the emergence of semiclassical spacetime from strongly interacting quantum dynamics.

\section*{Acknowledgments}
%We would like to thank 
The work of HN is supported in part by  CNPq grant 304583/2023-5 and FAPESP grant 2019/21281-4.
HN would also like to thank the ICTP-SAIFR for their support through FAPESP grant 2021/14335-0. 
ELG is supported by FAPESP grant 2024/13362-2.
JM and HJRVZ are supported in part by the “Quantum Technologies for Sustainable Development” grant 
from the National Institute for Theoretical and Computational Sciences of South Africa (NITHECS).
%CB is supported by the Oppenheimer Memorial Trust Research Fellowship and the Harry Crossley 
%Research Fellowship.

\appendix

\section{A single cycle sum}

\label{SingleCylceSum}

To get some handle on the expression in (\ref{AnAn}), consider the sum
$$f = n! \sum_{\tau \in S_n}   N^{2C(\tau)},$$
which involves a single sum over elements of $S_n$.  It is known that the permutations in $S_n$ can 
be arranged into conjugacy classes which can be labelled by Young diagrams.  The number of elements 
in a conjugacy class involving $k_j$ cycles of length $j$ satisfies is given by 
$$ \frac{n!}{\prod_{j=1}^n j^{k_j} k_j!}. $$
These are subject to the restriction that 
$$ \sum_{j=1}^n j k_j = n.$$  
The sum over elements of $S_n$ can be recast as 
\begin{eqnarray}
f & = & (n!)^2 \sum_{ \left\{  k_1, k_2, \cdots, k_n  \right\}    } \prod_{j=1}^n 
\left( \frac{N^2}{j} \right)^{k_j} \frac{1}{k_j !}    \nonumber \\
& = & (n!)^2 \sum_{ \left\{  k_1, k_2, \cdots, k_n  \right\}    } N^{2 K}\prod_{j=1}^n 
\left( \frac{1}{j} \right)^{k_j} \frac{1}{k_j !} \;,    \nonumber
\end{eqnarray}
where $$  K = \sum_{j} k_j. $$
In the above sum has two extremities\footnote{These correspond to a single row of length $n$ and a 
single column of length $n$.}, namely the conjugacy class of one-cycles and the conjugacy class 
of single cycles of length $n$.  These give the contributions
\begin{eqnarray}
f_{\left\{ n, 0, 0, \cdots, 0   \right\}} = (n!)^2 N^{2n} \frac{1}{n!} = n! N^{2n}   \nonumber \\
f_{\left\{ 0, 0, 0, \cdots, 1   \right\}} = (n!)^2 \frac{N^{2}}{n}  \frac{1}{1!} = n!(n-1)! N^{2}.    \nonumber
\end{eqnarray}
We can similarly compute the terms corresponding to various other partitions, e.g., 
$$ f_{\left\{ n-2, 1, 0, \cdots, 0   \right\}} = (n!) \frac{n(n-1)}{2} N^{2n-2}  $$
and 
$$ f_{\left\{ 1, 0, 0, \cdots, 1, 0   \right\}} = \frac{(n!)^2}{n-1} N^{4}  .$$

\section{Ricci tensor calculations}

Consider the more general ansatz
\be
ds^2=f(\vec{x},\vec{y})(dt+d\psi)^2+h_{ij}(\vec{x})dx^i dx^j+G_{ab}(\vec{y})dy^a dy^b\equiv G_{\mu\nu}
dx^\mu dx^\nu\;,
\ee
where $t,\psi$ are isometries ($\d_t G_{\mu\nu}=\d_\psi G_{\mu\nu}=0$) and, more specifically, 
\be
g_{tt}=f-h\;,\;\;\;
g_{\psi\psi}=f+g\;,\;\;\;
g_{t\psi}=f.
\ee

Then the inverse metric in the $(t,\psi)$ space is
\be
(g^{-1})_{t,\psi}=\frac{1}{|g|}\begin{pmatrix} f+g & -f\\ -f& f-h\end{pmatrix}\;,
\ee
and where $|g|$ is the determinant of the metric in the $(t,\psi)$ directions, 
\be
|g|=(f-h)(f+g)-f^2=-hg+f(g-h).
\ee
Denote by $\tilde\mu$ the coordinates other than $t,\psi$, so $x^{\tilde \mu}=(x^i,y^a)$, and their 
determinant by $|\tilde g|$, so 
\be
|G|=|g| \; |\tilde g|.
\ee
After a calculation, one finds that $f$ appears only in the following Christoffel symbols:
\bea
{\Gamma^t}_{t\tilde \mu}&=& \frac{1}{2|g|}(-g\d_{\tilde \mu} h+g\d_{\tilde \mu}f-f\d_{\tilde \mu}h)\;,\;\;
{\Gamma^t}_{\psi\tilde\mu}=\frac{1}{2|g|}(g\d_{\tilde\mu} f-f\d_{\tilde \mu}g)\cr
{\Gamma^\psi}_{\psi\tilde\mu}&=& \frac{1}{2|g|}(-h\d_{\tilde \mu}g +f \d_{\tilde \mu}g -h \d_{\tilde \mu} g)\;,\;\;
{\Gamma^\psi}_{t\tilde\mu}=\frac{1}{2|g|}(f\d_{\tilde \mu}h-h\d_{\tilde \mu}f)\cr
{\Gamma^{\tilde\mu}}_{tt}&=& -\frac{1}{2}g^{\tilde \mu \tilde \nu}\d_{\tilde \nu}(f-h)\;,\;\;
{\Gamma^{\tilde\mu}}_{\psi\psi}= -\frac{1}{2}g^{\tilde \mu \tilde \nu}\d_{\tilde \nu}(f+g)\;,\;\;
{\Gamma^{\tilde\mu}}_{t\psi}= -\frac{1}{2}g^{\tilde \mu \tilde \nu}\d_{\tilde \nu}(f).
\eea

Then, using the formula
\be
R_{\mu\nu}=\d_\rho {\Gamma^\rho}_{\mu\nu}-\frac{1}{2}\d_\mu \d_\nu \ln |G|
+\frac{1}{2}{\Gamma^\rho}_{\mu\nu}\d_\rho \ln |G| -{\Gamma^\rho}_{\mu\sigma}{\Gamma^\sigma}_{\rho\nu}\;,
\ee
we find that only $R_{tt}, R_{\psi\psi}$ and $R_{t\psi}$ contain $f$, namely as 
\bea
R_{tt}&=& -\frac{1}{2}\frac{1}{\sqrt{ |G|}}\d_{\tilde \mu }\sqrt{|G|} g^{\tilde\mu\tilde \nu}\d_{\tilde \nu}(f-h)
\cr
&&-\frac{1}{2|g|}g^{\tilde\mu\tilde \nu}[\d_{\tilde \nu }h(f\d_{\tilde \mu}h-2g \d_{\tilde \mu} f)+g\d_{\tilde \mu} h
\d_{\tilde \nu} h+(g-h)\d_{\tilde \mu} f \d_{\tilde \nu f}]+{\rm background}\cr
R_{\psi\psi}&=& -\frac{1}{2}\frac{1}{\sqrt{ |G|}}\d_{\tilde \mu }\sqrt{|G|} g^{\tilde\mu\tilde \nu}\d_{\tilde \nu}(f+g)
\cr
&&+\frac{1}{2|g|}g^{\tilde\mu\tilde \nu}[\d_{\tilde \nu }g(-f\d_{\tilde \mu}g+2h \d_{\tilde \mu} f)-h\d_{\tilde \mu} g
\d_{\tilde \nu} g+(g-h)\d_{\tilde \mu} f \d_{\tilde \nu f}]+{\rm background}\cr
R_{\psi t}&=& -\frac{1}{2}\frac{1}{\sqrt{ |G|}}\d_{\tilde \mu }\sqrt{|G|} g^{\tilde\mu\tilde \nu}\d_{\tilde \nu}(f)
\cr
&&+\frac{1}{2|g|}g^{\tilde\mu\tilde \nu}[\d_{\tilde \nu }g(-f\d_{\tilde \mu}g+2h \d_{\tilde \mu} f)-h\d_{\tilde \mu} g
\d_{\tilde \nu} g+(g-h)\d_{\tilde \mu} f \d_{\tilde \nu f}]+{\rm background}.\cr
&&
\eea
Actually, the other Riccis also contain $f$ generally, but under the condition we will impose, they will not 
anymore. 

Indeed, we see that the conditions for the Einstein equation to contain only terms linear in $f$ (plus background
terms, which vanish if the background is a solution of the Einstein's equations) are
\be
h=g\;,\;\;\; \d_{\tilde \mu}g=\d_{\tilde \mu }h=0\;,
\ee
or that $h$ and $g$ are (approximately, at least!) equal constants. Note, indeed, that there is no 
condition on the double derivatives: we can have $\d_{\tilde\mu}\d_{\tilde\nu}g\neq 0, \d_{\tilde \mu}\d_{\tilde
\nu}h \neq 0$. 

Then, near $r=0, \theta=0$, the above conditions are true, since 
\be
h=\cosh^2 r\;,\;\; g=\cos^2\theta\;,
\ee
so that 
\be
\left.\d_i h\right|_{r=0}=0\;,\;\;\;
\left.\d_a g\right|_{\theta=0}=0.
\ee
(as noted, the second derivatives at $r=0,\theta=0$ are nonzero, but we don't care about that!). Also note 
that, if $h-g$, then 
\be
|g|=-gh+f(g-h)=-gh\;,
\ee
is independent of $f$. 

\section{Additional Dyson-Schwinger equations}
Since we have an $\omega$ in $m=2\omega$, 
we denote the variable ("momentum") as $\omega_1$, and the other momenta as 
$\omega_2,\omega_3,\omega_4,\omega_5$), so the DS equation for the 2-point function of $\phi$ is
\bea
&&\tilde G_{2,\phi}(T,\omega_1)=\tilde G_{2,0,\phi}(T,\omega_1)-(\tilde g N)^2\tilde G_{2,0,\phi}
(T,\omega_1)\tilde G_{2,\phi}(T,\omega_1)\cr
&&\int_{-\infty}^{+\infty}
\frac{d\omega_2}{2\pi}\int_{-\infty}^{+\infty}\frac{d\omega_3}{2\pi}\int_{-\infty}^{+\infty}
\frac{d\omega_4}{2\pi}\int_{-\infty}^{+\infty}\frac{d\omega_5}{2\pi}
\tilde G_{6,X}(T;\omega_2,\omega_3,\omega_1-\omega_2-\omega_3; \omega_4,\omega_5,\omega_1-\omega_4-\omega_5)\cr
&&\label{SDphi}
\eea
(here we have assumed the 0-point function is zero). 

Indeed, in the logic of Schwinger-Dyson equation, we can: freely propagate; or freely propagate, 
then interact, then contract with the resulting 6-point function of $X$'s, etc.

For the $X$'s, we get the recursive Dyson-Schwinger equations: 
consider that the interaction with $\phi$, otherwise free, 
appears only via 4-point vertices, creating two extra legs for the $X$ $n$-point function, so 
\bea
&&\tilde G_{2,X}(\omega_1)= \tilde G_{2,X,0}(\omega_1)
-(\tilde gN)^2\tilde G_{2,X,0}(\omega_1)\tilde G_{2,X}(\omega_1)
\int_{-\infty}^{+\infty}\frac{d\omega_2}{2\pi}\int 
_{-\infty}^{+\infty}\frac{d\omega_3}{2\pi}\cr
&&\int_{-\infty}^{+\infty}\frac{d\omega_4}{2\pi}\tilde G_{2,0,\phi}(\omega_1-\omega_2-\omega_3)
\tilde G_{4,X}(\omega_2,\omega_3;\omega_4,\omega_2+\omega_3-\omega_4)\cr
&&\tilde G_{4,X}(\omega_1,\omega_2;\omega_3,\omega_1+\omega_2-\omega_3)=
\tilde G_{2,0,X}(\omega_1)\tilde G_{2,X}(\omega_2, \omega_3=\omega_2)+({\rm cyclic\;in\;1,2,3})\cr
&&
-(\tilde g N)^2G_{2,0,X}(\omega_1)\int_{-\infty}^{+\infty}\frac{d\omega_4}{2\pi}\int_{-\infty}^{+\infty}
\frac{d\omega_5}{2\pi}\int_{-\infty}^{+\infty}\frac{d\omega_6}{2\pi}
\int_{-\infty}^{+\infty}\frac{d\omega_7}{2\pi}\cr
&&\tilde G_{2,0,\phi}(\omega_1-\omega_4-\omega_5)
\tilde G_{6,X}(\omega_2,\omega_4,\omega_5;\omega_6,\omega_7,\omega_2
+\omega_4+\omega_5-\omega_6-\omega_7)\cr
&&
\tilde G_{4,X}(\omega_2+\omega_4+\omega_5-\omega_6-\omega_7,\omega_1
-\omega_4-\omega_5+\omega_6+\omega_7;
\omega_3,\omega_1+\omega_2-\omega_3)\cr
&&\tilde G_{6,X}(\omega_1,\omega_2,\omega_3;\omega_4,\omega_5,\sum_i\omega_i)
=\tilde G_{2,X,0}(\omega_1)\tilde G_{4,X}
(\omega_2,\omega_3;\omega_4,\omega_5=...)+({\rm cyclic\;in\;1,...,5})\cr
&&-(\tilde gN)^2G_{2,0,X}(\omega_1)\int_{-\infty}^{+\infty}\frac{d\omega_6}{2\pi}\int_{-\infty}^{+\infty}
\frac{d\omega_7}{2\pi}\int_{-\infty}^{+\infty}\frac{d\omega_8}{2\pi}\int_{-\infty}^{+\infty}\frac{d\omega_9}{2\pi}
\int_{-\infty}^{+\infty}\frac{d\omega_{10}}{2\pi}\cr
&&\tilde G_{2,0,\phi}(\omega_1-\omega_6-\omega_7)
\tilde G_{8,X}(\omega_2,\omega_3,\omega_6,\omega_7;\omega_8,\omega_9,\omega_{10},\cr
&&\omega_2+\omega_3+\omega_6+\omega_7-\omega_8-\omega_9-\omega_{10})\cr
&&\tilde G_{6,X}(\omega_2+\omega_3+\omega_6+\omega_7-\omega_8-\omega_9-\omega_{10},\omega_{10},\cr
&&\omega_1-\omega_6-\omega_7+\omega_8+\omega_9;
\omega_4\omega_5,\omega_1+\omega_2+\omega_3-\omega_4-\omega_5)\;,
\eea
etc.  In the above equations, we have 
not put the temperature dependence, but it should be there. 

So it is an infinite series, that in principle would need to be solved all at once, but we must cut the 
series at some point (by considering the top $n$-point functions as free) and solve for the lower point functions.
We also cannot ignore the temperature anymore for 
neither the $X$ nor the $\phi$ propagators, since the temperature is higher than both masses. 

We the solve iteratively the equations, by putting 
$G_{4,X}=G_{4,X,0}$, $G_{6,X}=G_{6,X,0}$ and $G_{8,X}=G_{8,X,0}$
(so with only free propagators) on the right-hand side of the $\tilde G_{6,X}$ equation, 
and finding $G_{6,X}^{(1)}$ (at iteration step 1) from the left-hand side. 

At most what we can say is that we consider also $\tilde g$ to be small, and solve the Schwinger-Dyson equation 
for $\phi$ by replacing $\tilde G_{6,X}$ with the product of 3 free $\tilde G_{2,0,X}$'s, and obtain 
\bea
\tilde G_{2,\phi}(T,\omega_1)&=&\tilde G_{2,0,\phi}(T,\omega_1)\left[1+(\tilde g N)^2G_{2,0,\phi}(T,\omega_1)
\int_{-\infty}^{+\infty}\frac{d\omega_2}{2\pi}
\int_{-\infty}^{+\infty}\frac{d\omega_3}{2\pi}\right.\cr
&&\left.\tilde G_{2,0,X}(T,\omega_2)\tilde G_{2,0,X}(T,\omega_3)
\tilde G_{2,0,X}(T,\omega_1-\omega_2-\omega_3)\right]^{-1}.
\eea

In reality, this gives the solution for $\tilde G_{2,\phi}(T,\omega_1)$ at iteration step 1. 

But it is not clear how good  this approximation is (or when it is good).
As before (since $m^2=4\omega^2$),
\bea
\tilde G_{2,0,X}(T;\omega_1)&=&\frac{i}{1-e^{-(2\omega)T}}\left(\frac{1}{\omega_1^2-(2\omega)^2+i\epsilon}
-\frac{e^{-(2\omega)/T}}{\omega_1^2-(2\omega)^2-i\epsilon}\right)\cr
&\simeq& \frac{i}{2\omega/T}\left(\frac{1}{\omega_1^2-(2\omega)^2+i\epsilon}
-\frac{1-(2\omega)/T}{\omega_1^2-(2\omega)^2
-i\epsilon}\right).
\eea

The general solution is actually 
\bea
\tilde G_{2,\phi}(T,\omega_1)&=&\tilde G_{2,0,\phi}(T,\omega_1)\left[1+\right.\cr
&&\left.+(\tilde g N)^2G_{2,0,\phi}(T,\omega_1)
\int_{-\infty}^{+\infty}\frac{d\omega_2}{2\pi}
\int_{-\infty}^{+\infty}\frac{d\omega_3}{2\pi}
\int_{-\infty}^{+\infty}\frac{d\omega_4}{2\pi}
\int_{-\infty}^{+\infty}\frac{d\omega_5}{2\pi}
\right.\cr
&&\left.\tilde G_{6,X}(T;\omega_2,\omega_3,\omega_1-\omega_2
-\omega_3;\omega_4,\omega_5,\omega_1-\omega_4-
\omega_5)\right]^{-1}.
\eea

Putting the solution for $G_{6,X}$ at iteration step 1 in the right-hand side, 
we obtain on the left-hand side the solution for $G_{2,\phi}(T,\omega_1)$ at iteration step 2. 

However, a better way to obtain Dyson-Schwinger equations was shown in the text.

%{\em Eric, please change/add to this according to what you did}.

\bibliography{KComplexityBH}

\bibliographystyle{utphys}

\end{document}